\documentclass[11pt,a4paper]{article}
\usepackage{jheppub}

\usepackage[utf8]{inputenc}
\usepackage{graphicx}
\usepackage{amsmath}
\allowdisplaybreaks[2]

\newcommand{\chpt}{ChPT}
\newcommand{\figWidth}{0.464}

\preprint{\begin{minipage}{4cm}\begin{flushright}
LU TP 14-03\\
Revised March 2014
\end{flushright}
\end{minipage}
}

\title{Masses, Decay Constants and Electromagnetic Form-factors
 with Twisted Boundary Conditions}

\author[a]{Johan Bijnens}
\author[a]{and Johan Relefors}
\affiliation[a]{Department of Astronomy and Theoretical Physics,
Lund University,\\
S\"olvegatan 14A, SE - 223 62 Lund, Sweden}
\emailAdd{bijnens@thep.lu.se}
\emailAdd{Johan.Relefors@thep.lu.se}

\abstract{
Using Chiral Perturbation Theory at one-loop we analyze
the consequences of twisted boundary conditions. We point out that due to the
broken Lorentz and reflection symmetry a number of unexpected terms
show up in the expressions. We explicitly discuss the pseudo-scalar octet
masses,
axial-vector and pseudo-scalar decay constants and electromagnetic form-factors.
We show how the Ward identities are satisfied using the momentum dependent
masses and the non-zero vacuum-expectation-values values for the electromagnetic
(vector) currents. Explicit expressions at one-loop are provided and an appendix
discusses the needed one-loop twisted finite volume integrals.
}


\begin{document}

\maketitle

\section{Introduction}
\label{sec:introduction}

Lattice QCD calculations of hadronic quantities of necessity
happen in a finite physical volume. In a box with periodic boundary
conditions this leads to spatial momentum components $p^i =(2\pi/L)n_i$
which even for a large 4~fm lattice gives a minimum spatial momentum of
about 300~MeV. In order to access smaller spatial momenta it has been suggested
to use twisted boundary conditions \cite{Bedaque,twisted,twisted2}.
This allows for more momenta to be sampled.
Some early numerical tests were performed in \cite{UKQCD1}.

It is well known that in a finite box Lorentz invariance is broken by the
boundary conditions. In particular, the spatial part of the symmetry group
becomes the cubic group in case of periodic boundary conditions.
Imposing twisted boundary conditions on a field $\phi$ in some spatial
directions $i$ via
\begin{align}
\label{deftwist}
\phi(x^i+L) = e^{i\theta_i L}\phi(x^i)
\end{align}
breaks the cubic symmetry down even further. In particular, reflection symmetry,
$x^i\to -x^i$ in the $i$-direction is broken by (\ref{deftwist}).

In this paper we analyze the consequences of this for a number of quantities
in Chiral Perturbation Theory (\chpt).
In \cite{twisted} \chpt\ for twisted boundary conditions was developed
and they showed that finite volume corrections remain exponentially suppressed
for large volumes. We use their method for masses, pseudo-scalar and
axial-vector decay constants, the vector two-point function and electromagnetic
form-factors.
We have different expressions than those given in \cite{twisted}, the
precise relation is discussed in more detail in Sect.~\ref{sec:comparison}.

In general, form-factors and correlators can also have a much more general
structure and this has consequences for the Ward identities. We discuss
three examples of this. Another result is that vector currents get a
vacuum-expectation-value (VEV), which leads to non-transverse vector
two-point functions. The main goal of our paper is to study all this
at one-loop order in \chpt.

Sect.~\ref{sec:chpt} gives the lowest order Lagrangian in \chpt\ and
defines a few other pieces of notation. We introduce
twisted boundary conditions in Sect.~\ref{sec:finitevolumeandtwist}.
The more technical derivation of the needed one-loop integrals is
given in App.~\ref{app:Integrals}. As a first application we calculate the
vacuum expectation value of vector currents and the two-point functions.
We show how they do satisfy the Ward identities at finite volume.
We find, in agreement with \cite{Aubin}, that the two-point function is not
transverse. The next two sections contain the results for the meson masses
and the axial-vector and pseudo-scalar decay constants. Here again we see
the occurrence of extra terms. The axial-vector matrix elements is not just
described by the decay constant but there are other terms.
The pseudo-scalar decay constants at infinite volume were not published earlier
so we have included those expressions as well. We have explicitly checked that
the Ward identities relating the axial-vector
and pseudo-scalar matrix elements are satisfied. The extra terms
in the axial-vector matrix element are needed to achieve this. We also add
the mixed matrix elements due to the fact that the twisted boundary conditions
break isospin. Numerical results are presented for all masses
and the charged meson axial-vector decay constants.

Sect.~\ref{sec:formfactor} discusses the pion electromagnetic form-factor
and related quantities. We show once more how finite volume and twisting
allow for extra form-factors and have checked that with the inclusion
of these the Ward identities are satisfied. We study in detail the finite volume
corrections from the isospin current matrix element
$\left<\pi^0(p^\prime)|\bar{d}\gamma_\mu u |\pi^+(p)\right>$ which is used
in lattice QCD to obtain information on the pion radius. We find that the
corrections due to twisting can be sizable. Our main conclusions are
summarized in Sect.~\ref{sec:conclusions}.

After finishing this work we became aware of the work in \cite{JT} where
a number of the issues we discuss here were raised as well.
The discussion there is in two-flavour theory but also includes partial
twisting. We discuss the relation with our work in Sect.~\ref{sec:comparison}.

\section{Chiral Perturbation Theory}
\label{sec:chpt}

\chpt\ is the effective field theory describing low energy QCD as an expansion
in masses and momenta \cite{Weinberg,GL1,GL2}. Finite volume \chpt\ was
introduced in \cite{GL4}. In this paper we work in the isospin limit for
quark masses, i.e.
$m_u=m_d=\hat m$, with
three quark flavours. Results for two-quark flavours are obtained by simply
dropping the integrals involving kaons and eta and replacing $F_0,B_0$ by $F,B$.
We perform the calculations to next-to-leading order (NLO),
or $\mathcal{O}(p^4)$. The Lagrangian to NLO is
\begin{align}
\mathcal{L} = \mathcal{L}_2 + \mathcal{L}_4,
\end{align}
where $\mathcal{L}_{2n}$ is the $\mathcal{O}(p^{2n})$ Lagrangian. 
For the mesonic fields we use the exponential representation
\begin{align}
  U = e^{i\sqrt{2}M/F_0}
\quad
\text{with}
\quad
  M =
  \left(\begin{matrix}
    \frac{1}{\sqrt{2}}\pi^0+\frac{1}{\sqrt{6}}\eta & \pi^+ & K^+ \\
    \pi^- & -\frac{1}{\sqrt{2}}\pi^0+\frac{1}{\sqrt{6}}\eta & K^0 \\
    K^- & \bar{K}^0 & -\frac{2}{\sqrt{6}}\eta
  \end{matrix}\right).
\end{align}
We use the external field method \cite{GL1,GL2} to incorporate
electromagnetism, quark masses as well as couplings to other quark-antiquark
operators.
To do this we introduce the field $\chi$ and the covariant derivative
\begin{align}
   \chi = 2B_0(s+ip),\quad
 D_\mu U = \partial_\mu U - ir_\mu U + iUl_\mu.
\end{align}
$r_\mu$, $l_\mu$, $s$ and $p$ are the external fields. 
Electromagnetism is included by setting
\begin{align}
  l_\mu = eA_\mu Q, \, r_\mu = eA_\mu Q,
\end{align}
where $e$ is the electron charge, $A_\mu$ is the photon field and
$Q=\text{diag}(2/3,-1/3,-1/3)$. Masses are included by setting
$s=\mathcal{M} = \text{diag}(\hat m,\hat m,m_s)$ where $\hat m = (m_u+m_d)/2$.

With these definitions the lowest order Lagrangian $\mathcal{L}_2$ is
\begin{align}
\mathcal{L}_2 = \frac{F_0^2}{4}
\big<D_\mu U D^\mu U^\dagger
 + \chi U^\dagger + U\chi^\dagger\big>
\end{align}
where the angular brackets denotes trace over flavour indices. The expression
for $\mathcal{L}_4$ can be found in for example \cite{GL1}.

One problem at finite volume is the definition of asymptotic states, which we
need to define the wave function renormalization and matrix elements.
We assume the temporal direction to be infinite in extent
and use the LSZ theorem to obtain the needed 
wave function renormalization by keeping the spatial momentum constant
and taking the limit in $(p^0)^2$ to $p^2=m^2$.
We stick here to states with at most one incoming and outgoing particle
so this is sufficient.
Note that since Lorentz symmetry is broken the masses are different for
the same particle with different spatial momenta.

We will not present the infinite volume expressions but only the corrections
at finite volume using the quantity
\begin{equation}
 \label{eq:presenting}
 \Delta^V X = X(V) - X(\infty),
\end{equation}
where $X$ is the object under discussion.

\section{Finite volume with a twist}
\label{sec:finitevolumeandtwist}

Periodic boundary conditions on a finite volume implies that momenta become
quantized. Adding a phase factor at the boundary shifts these discrete momenta.
To see this, we impose for a field in one dimension at a fixed time
\begin{equation} 
\label{eq:twist}
  \psi(x+L) = e^{i\theta}\psi(x),
\end{equation}
where $L$ is the length of the dimension and $\theta$ is the twist angle.
Developing both sides in a Fourier series we get
\begin{equation}
  \sum_k\hat{\psi}_ke^{ik(x+L)} =
 \sum_k\hat{\psi}e^{ikx}e^{i \theta}
 \Rightarrow k = \frac{2\pi}{L}n + \frac{\theta}{L}, \, n\in \mathbb{Z}.
\end{equation}
The effect on anti-particles follows from the complex conjugate
of (\ref{eq:twist}); momenta are shifted in the opposite direction. 
It is possible to have different twists for different flavours and also
different twists in different directions.

We impose now a condition like (\ref{eq:twist}) on each quark field $q$ in
each spatial direction $i$
\begin{equation} 
\label{eq:twistq}
  q(x^i+L) = e^{i\theta^i_q}q(x^i),
\end{equation}
and collect the angles $\theta^i_q$ in a three vector $\vec\theta_q$
and a four-vector $\theta_q=(0,\vec\theta_q)$. The twist-angle vector for the
anti-quark is minus the one for the quarks.
For a meson field of flavour structure $\bar q' q$ this leads to a twisted
boundary condition in direction $i$
\begin{equation}
\label{eq:twistqqp}
\phi_{\bar q' q}(x^i+L) = e^{i (\theta_q^i-\theta_{q'}^i)}\phi_{\bar q' q}(x^i)\,.
\end{equation}
We introduce the meson twist angle vector $\theta_{\phi}$ in the same way as
above and we will use the conventional $\pi^\pm,\ldots$ for labeling them.. 
Note that flavour diagonal mesons are unaffected by twisted boundary conditions.
A consequence of the boundary conditions (\ref{eq:twistqqp}) is that
charge conjugation is broken since $\phi_{\bar q q^\prime}$
and $\phi_{\bar q^\prime q}$ have opposite twist. A particle with spatial momentum
$\vec p$ corresponds to an anti-particle with momentum $-\vec p$.

In terms of loop integrals over the momentum of a meson
$M$ this means that we have to replace the infinite volume
integral by a sum over the three spatial momenta and an integral
over the remaining dimensions
\begin{equation}
\label{defintV}
  \int \frac{d^d k_M}{(2\pi)^2} \rightarrow
\int_ V \frac{d^d k}{(2\pi)^d} \equiv
 \int \frac{d^{d-3}k}{(2\pi)^{d-3}} \frac{1}{L^3}
\sum_{\begin{smallmatrix}\vec n\in \mathbb{Z}^3\\
          \vec k =(2\pi\vec n+\vec\theta_M)/L\end{smallmatrix}}\,. 
\end{equation}
It is explained in \cite{twisted} how this ends up with
the correct allowed momenta for each propagator in a loop.
The allowed momenta $\vec k =(2\pi\vec n+\vec\theta_M)/L$ are not  symmetric
around zero and thus reflection symmetry is broken.
An immediate consequence is that
\begin{equation}
\label{eq:tadpolenonzero}
  \int_V \frac{d^d k}{(2\pi)^2}\frac{k^\mu}{k^2-m^2} \neq 0\,.
\end{equation}
Note also that a meson and its anti-meson carry different momenta and
it is therefore important to keep track of which one is in a loop, as well
as to be careful with using charge conjugation. The twist angles
also bring in another source of explicit flavour symmetry breaking.

The one-loop integrals needed are worked out using the methods of
\cite{Becirevic,sunsetfiniteV} and presented in detail in
App.~\ref{app:Integrals}. The notation we use indicates the mass
of the particle but implies also the corresponding twist vector in the
expressions.

\section{Vector vacuum-expectation-value and two-point function}
\label{sec:vector}

Because of (\ref{eq:tadpolenonzero}) the vacuum-expectation-value of
a vector-current is non-zero and we obtain
\begin{align}
\big< \bar u\gamma_\mu u \big> &= -2 A^V_\mu(m_{\pi^+}^2)-2 A^V_\mu(m_{K^+}^2)
\notag\\
\big< \bar d\gamma_\mu d \big> &=  2 A^V_\mu(m_{\pi^+}^2)-2 A^V_\mu(m_{K^0}^2)
\notag\\
\big< \bar s\gamma_\mu s \big> &=  2 A^V_\mu(m_{K^+}^2)+2 A^V_\mu(m_{K^0}^2)
\notag\\
\big< j^{em}_\mu \big> &= -2 A^V_\mu(m_{\pi^+}^2)-2 A^V_\mu(m_{K^+}^2)\,.
\end{align}
We used here that $\theta_{\pi^-}=-\theta_{\pi^+}$,
$\theta_{K^+}=-\theta_{K^-}$, $\theta_{K^0}=-\theta_{\overline{K}^0}$
and $\theta_{\pi^0}=\theta_{\eta}=0$. This non-zero result can be understood
better if we look at the alternative way of including twisting
in \chpt\ \cite{twisted}. The twisted boundary conditions can be
removed by a field redefinition. However, then we
get a non-zero external vector field which can be seen as a constant
background field. Charged particle-anti-particle vacuum fluctuations
are affected by this background field thus giving rise to a non-zero
current even in the vacuum.

The two-point function of a current $j^\mu$ is defined as
\begin{equation}
\Pi_{\mu\nu}^{a}(q) \equiv i\int d^4x e^{iq\cdot x}
\big<T(j_\mu^a(x)j_\nu^{a\dagger}(0))\big>
\,.
\end{equation}
The current $j^{\pi^+}_\mu = \bar d\gamma_\mu u$ satisfies the Ward identity.
\begin{equation}
\label{eq:VVWard}
\partial^\mu\big<T(j^{\pi^+}_\mu(x)j^{\pi^-}_\nu(0))\big>=
\delta^{(4)}(x)\big<\bar d\gamma_\nu d-\bar u\gamma_\nu u\big>
\,.
\end{equation}
We used here that $m_u=m_d$ with the usual techniques to derive
Ward identities.
A consequence is that with twisted boundary conditions the vector
two-point function is no longer transverse. However,
flavour diagonal currents like the
electromagnetic one remain transverse. This does not mean that they
are proportional to $q_\mu q_\nu-q^2 g_{\mu\nu}$ since Lorentz symmetry is
broken. A more thorough discussion at the quark level and estimates using
lattice calculations can be found in \cite{Aubin}.

The infinite volume expressions we obtain agree with those
of \cite{ABT}. The finite-volume corrections for the $\bar d\gamma_\mu u$ and
electromagnetic current are
\begin{align}
\Delta^V\Pi^{\pi^+}_{\mu\nu}(q)&= 2\widetilde\Pi_{\mu\nu}(m_{\pi^+}^2,m_{\pi^0}^2,q)
 +\widetilde\Pi_{\mu\nu}(m_{K^+}^2,m_{\overline K^0}^2,q)\,,
\nonumber\\
\Delta^V\Pi^{em}_{\mu\nu}(q)&=  \widetilde\Pi_{\mu\nu}(m_{\pi^+}^2,m_{\pi^-}^2,q)
 +\widetilde\Pi_{\mu\nu}(m_{K^+}^2,m_{K^-}^2,q)\,,
\nonumber\\
\widetilde\Pi_{\mu\nu}(m_1^2,m_2^2,q)&=
g_{\mu\nu}\left( 4B_{22}^V(m_1^2,m_2^2,q)-A^V(m_1^2)-A^V(m_2^2)\right)
\nonumber\\ &
+q_\mu q_\nu\left( 4 B_{21}^V(m_1^2,m_2^2,q^2)-4 B_1^V(m_1^2,m_2^2,q^2)
  +B^V(m_1^2,m_2^2,q^2)\right)
\nonumber\\ &
+(q_\mu g_\nu^\alpha+q_\nu g_\mu^\alpha)(-2) B_{2\alpha}^V(m_1^2,m_2^2,q)
+4 B_{23\mu\nu}^V(m_1^2,m_2^2,q)\,.
\end{align}
Using the relations (\ref{eq:Brelations}) it can be checked that 
the consequences of (\ref{eq:VVWard}), namely
$q^\mu\Pi^{\pi^+}_{\mu\nu}= \big<\bar u\gamma_\mu u-\bar d\gamma_\mu d\big>$
and $q^\mu\Pi^{em}_{\mu\nu} = 0$ are satisfied.

We do not present numerical results here, the values of the vacuum expectation
value are small compared to $\big< \overline u u\big>$.

\section{Meson masses}
\label{sec:mass}

We define the mass here as the pole of the full propagator at fixed
spatial momentum $\vec p$. $\vec p$ should be such that it satisfies
the twisted boundary condition for the field under consideration.
Lorentz and charge conjugation invariance are broken
by the twisted boundary conditions. This leads to a mass that
depends on all components of the spatial momentum $\vec p$.
An anti-particle with spatial
momentum $-\vec p$ has the same mass as the corresponding particle with spatial
momentum $\vec p$.

The analytical results for the mass correction in terms of the integrals
defined in App.~\ref{app:Integrals} are
\begin{align}
\label{resultmasses}
\Delta^V\! m_{\pi^\pm}^2 &= 
\frac{\pm p^\mu}{F_0^2}
    [- 2 A^V_\mu(m_{\pi^+}^2)-  A^V_\mu(m_{K^+}^2) +  A^V_\mu(m_{K^0}^2)]
\nonumber\\
&+\frac{m_\pi^2}{F_0^2}
\left( - \frac{1}{2} A^V(m_{\pi^0}^2) + \frac{1}{6}A^V(m_{\eta}^2)\right)\,,
\nonumber\\
\Delta^V\! m_{\pi^0}^2 &= \frac{m_\pi^2}{F_0^{2}}
\left(
  - A^V(m_{\pi^+}^2) + \frac{1}{2}A^V(m_{\pi^0}^2) + \frac{1}{6}A^V(m_{\eta}^2)
\right)\,,
\nonumber\\
\Delta^V\! m_{K^\pm}^2&= \pm\frac{p^\mu}{F_0^2}
      [  - A^V_\mu(m_{\pi^+}^2) - 2 A^V_\mu(m_{K^+}^2) - A^V_\mu(m_{K^0}^2)]
 - \frac{m_K^2}{F_0^2}\frac{1}{3}A^V(m_{\eta}^2)\,,
\nonumber\\
\Delta^V\!m_{K^0(\overline{K}^0)}^2 &= +(-) \frac{p^\mu}{F_0^2}
 [ A^V_\mu(m_{\pi^+}^2) - A^V_\mu(m_{K^+}^2) - 2 A^V_\mu(m_{K^0}^2)]
  -\frac{m_K^2}{F_0^2} \frac{1}{3} A^V(m_{\eta}^2)\,,
\nonumber\\
\Delta^V\! m_{\eta}^2 &=  
   -\frac{m_K^2}{F_0^2}\frac{2}{3} (A^V(m_{K^+}^2)+A^V(m_{K^0}^2))
   +\frac{m_\eta^2}{F_0^2}\frac{2}{3}A^V(m_{\eta}^2)\,,  
\nonumber\\
&+\frac{m_{\pi}^2}{F_0^2}\frac{1}{6}(2A^V(m_{\pi^+}^2) +  A^V(m_{\pi^0}^2) - A^V(m_{\eta}^2))\,.
\end{align}
The notation $K^0(\overline K^0)$ and $+(-)$ means $+$ for $K^0$ and $-$ for
$\overline K^0$.
We agree with the infinite volume
expressions of \cite{GL2} and the known untwisted finite-volume corrections
\cite{GL4,Becirevic}. The relation to the results
in \cite{twisted,JT} is discussed in Sect.~\ref{sec:comparison}.

In (\ref{resultmasses}) the masses $m_\pi^2$, $m_K^2$ and $m_\eta^2$
can be replaced by the physical masses with or without finite volume
correction, or lowest order masses.
The differences are higher order. The same comment applies to $F_0$ in
(\ref{resultmasses}). The masses in the loop functions $A^V$
are written as the physical masses. The notation
$A^V(m_M^2)$ with $M$ the meson includes includes the dependence on
$\theta_M$. We keep for example $\pi^+$ and $\pi^0$ as notation
even if they have the same infinite volume and lowest order mass,
since $\theta_{\pi^+}$ and $\theta_{\pi^0}$ are different.

Note that in the case where $\vec{p} = \vec{\theta}/L$ the different signs
for $A^V_\mu$ between particle and anti-particle will be canceled by
the sign difference in $\vec p$ originating from opposite twist angles. 
The same cancellation occurs for the higher momentum states if the change
 $2\pi\vec n/L\rightarrow -2\pi\vec n/L$ is taken.
This is consistent with the fact that charge conjugation should be defined with a change
of sign in momentum, as discussed above.

The twisted boundary conditions do break isospin and thus induce $\pi^0$-$\eta$
mixing. This only affects the masses at next-to-next-to-leading-order (NNLO),
i.e. higher order than NLO.
The derivation follows the arguments as given in
Sect.~2.1 in \cite{ABT2} .

\begin{figure}[tp]
\begin{center}
\includegraphics[width=\figWidth\textwidth]{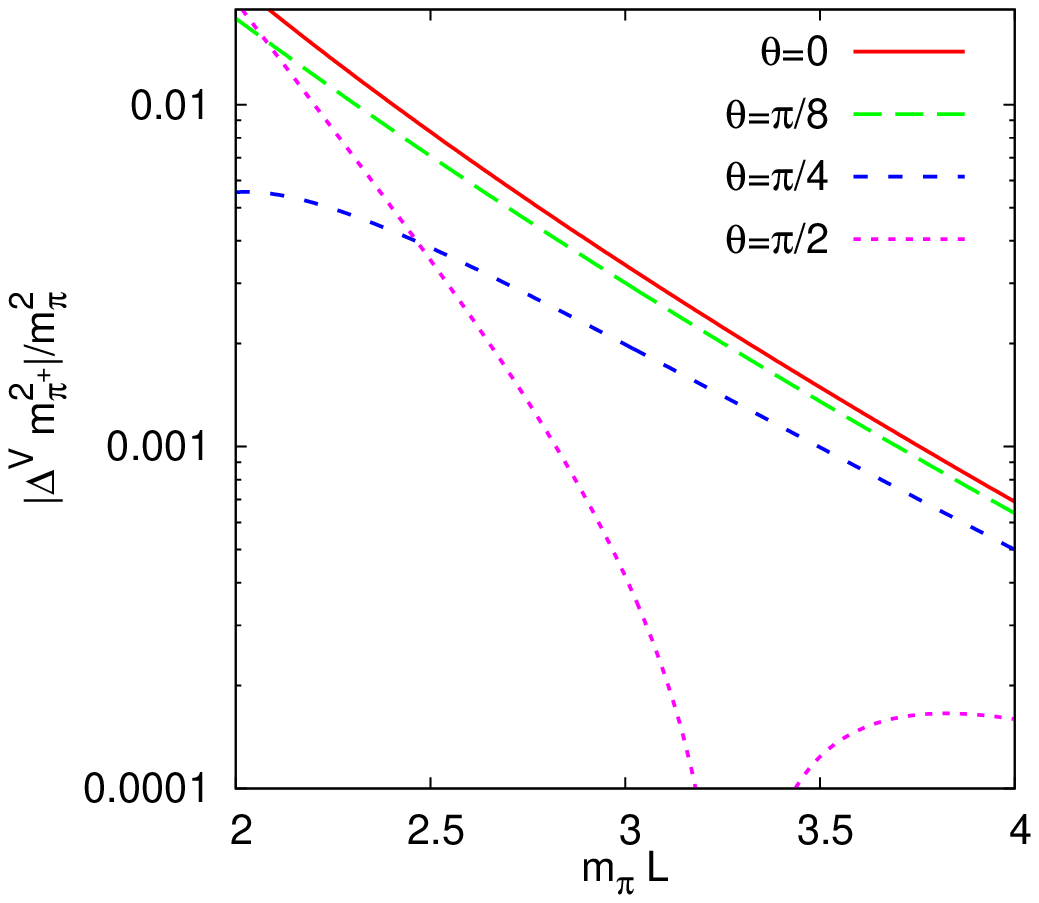}
\hspace*{0.05\textwidth}
\includegraphics[width=\figWidth\textwidth]{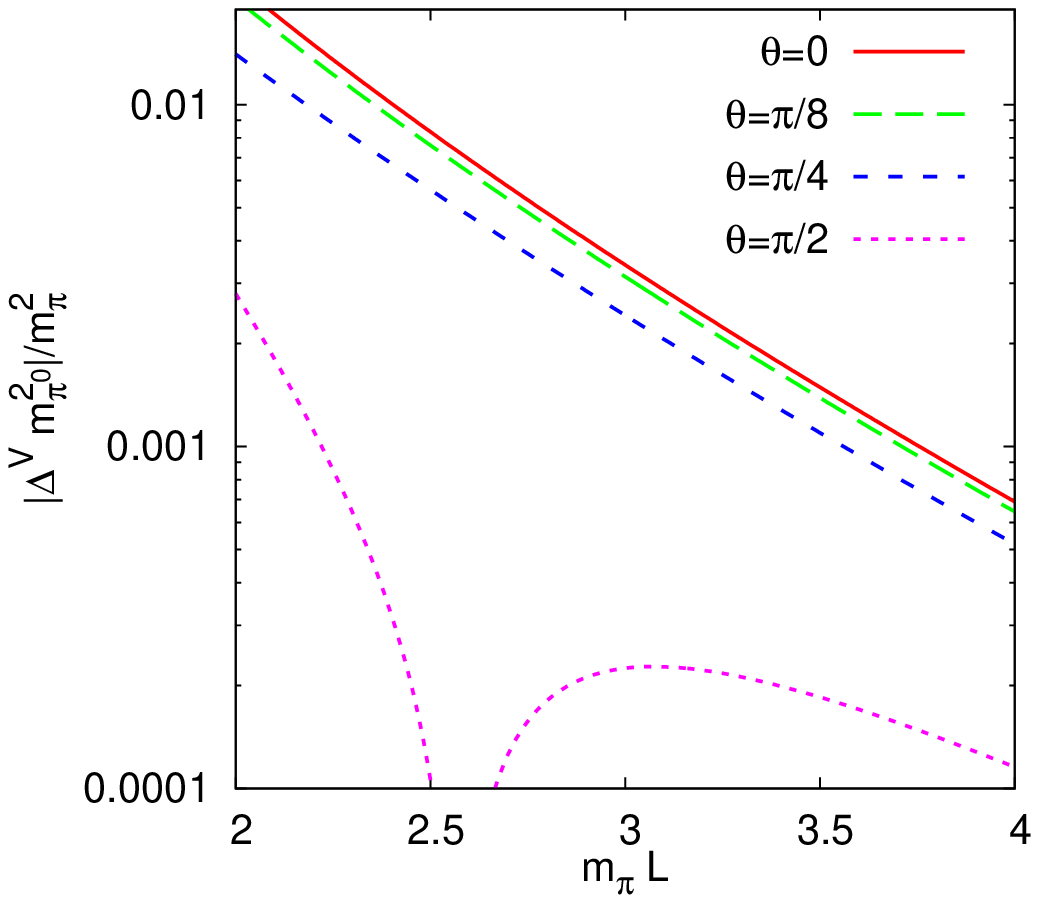}
\\
\includegraphics[width=\figWidth\textwidth]{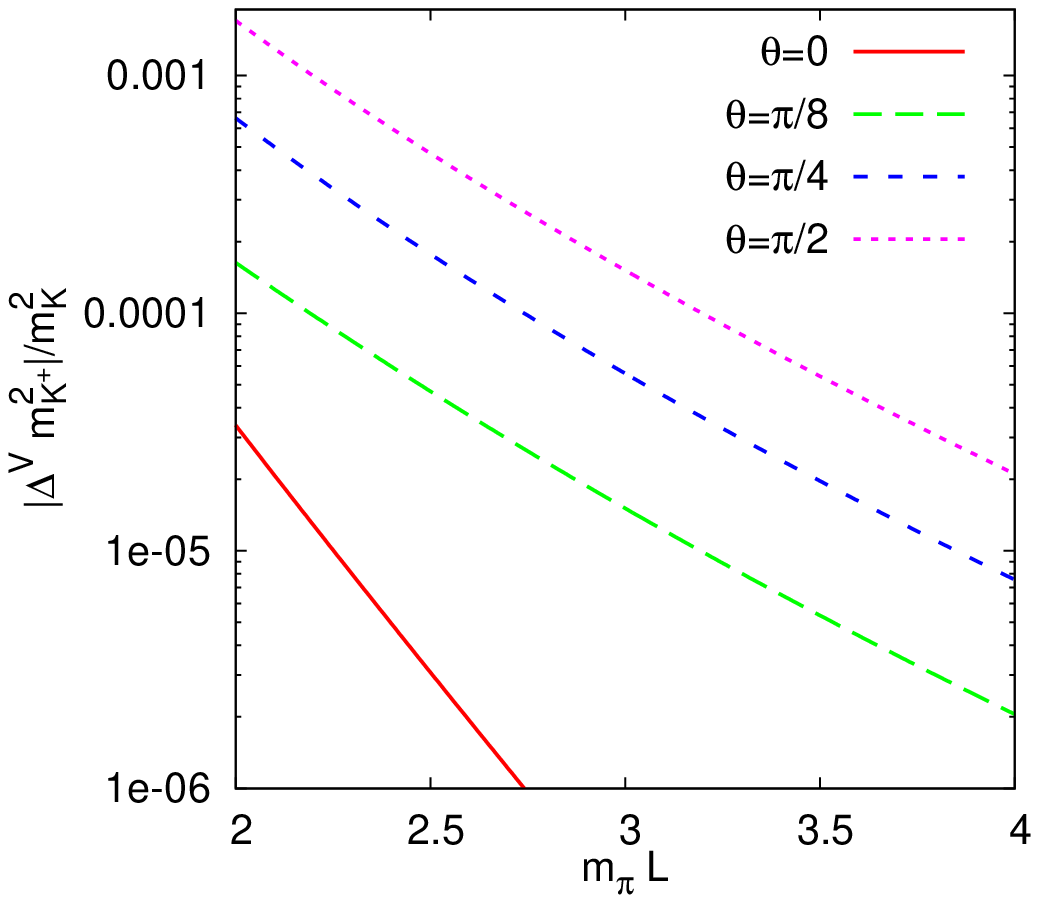}
\hspace*{0.05\textwidth}
\includegraphics[width=\figWidth\textwidth]{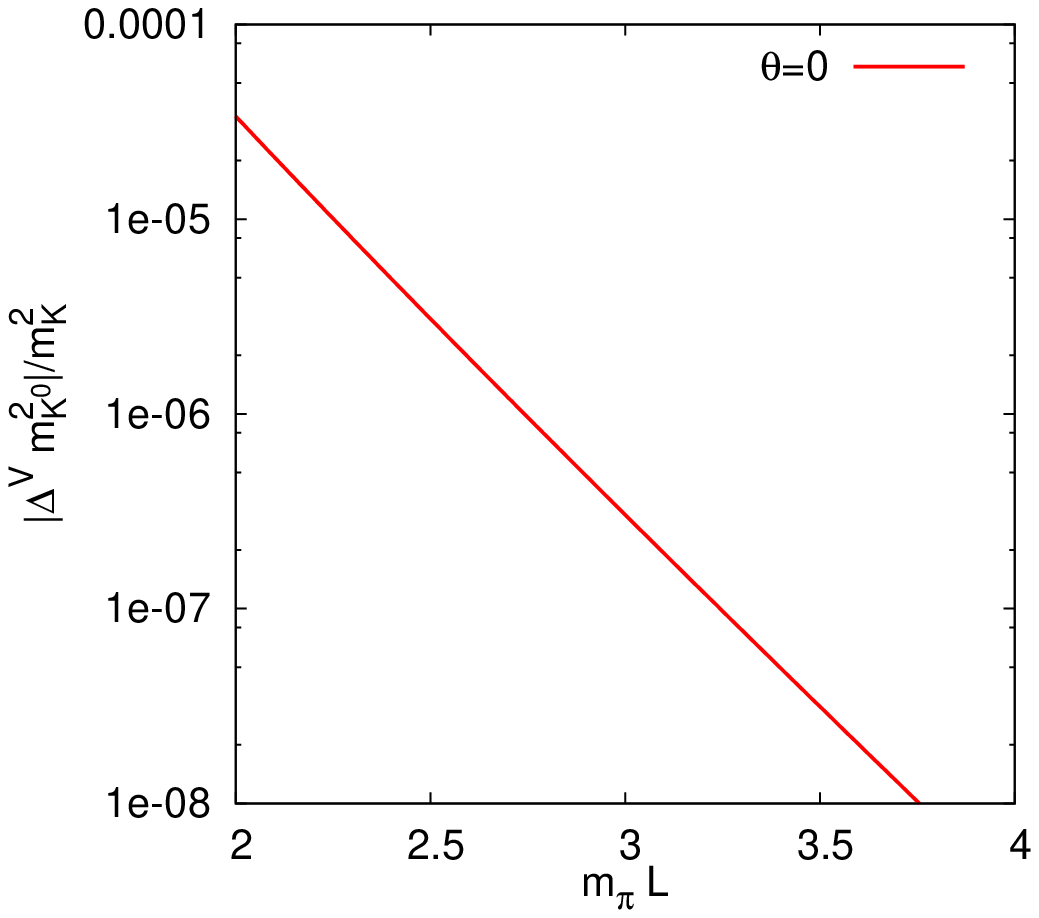}
\\
\includegraphics[width=\figWidth\textwidth]{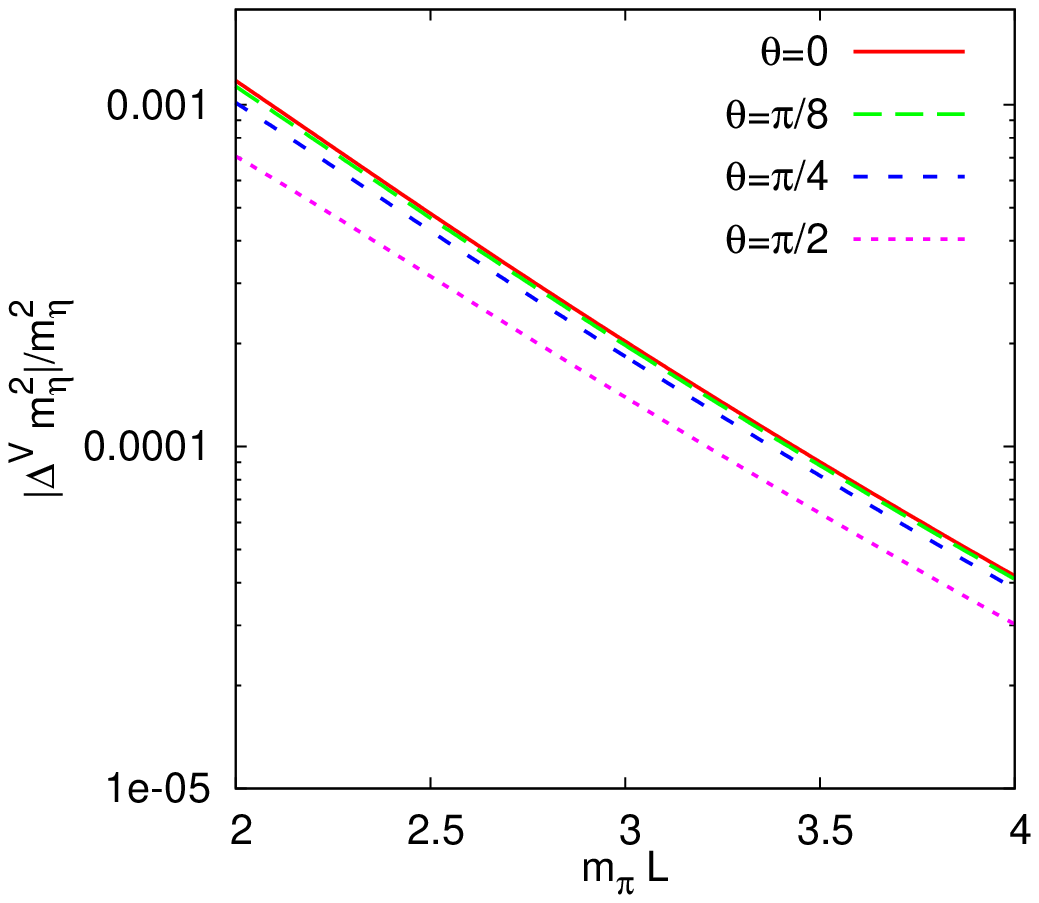}
\end{center}
\caption{\label{fig:masses}
Absolute value of the relative finite volume correction to the masses of
the light pseudo-scalar mesons
as a function of the box size for various twist angles.
The twist is for all cases on the up quark. The input values are
specified in (\ref{input1}) and (\ref{input2}).
The dip in the top two plots is where the correction goes through zero}
\end{figure}

We now show the volume and twist angle dependence for the case with
\begin{equation}
\label{input1}
m_\pi = 139.5~\mathrm{MeV}\,,\quad m_K = 495~\mathrm{MeV}\,,
\quad m_\eta^2 = \frac{4}{3} m_K^2-\frac{1}{3}m_\pi^2,\quad
F_\pi = 92.2~\mathrm{MeV}\,.
\end{equation}
We have used these masses in the one-loop expressions as well
as the value of $F_\pi$ for $F_0$ in the expressions.
We show results for several values of the twist angle $\theta$ with
\begin{equation}
\label{input2}
\vec\theta_u=(\theta,0,0)\,,\quad \vec\theta_d = \vec\theta_s = 0\,.
\end{equation}
Note that this implies that for $\pi^+$ and $K^+$ there is a non-zero 
spatial momentum $\vec p= \vec \theta_u / L$, while $\vec p$ vanishes for
$\pi^0$, $K^0$ and $\eta$.
As can be seen in Fig.~\ref{fig:masses}, the finite volume correction has a
sizable dependence on the twist-angle. The correction for the $K^0$
does not depend on the twist angle here, since for the choice of angles in
(\ref{input2}) there is only the $\eta$-loop contribution due to
$\vec p_{K^0}=0$. The relative correction to the kaon and eta masses remains
small while for $\pi^+$ and $\pi^0$ it can become in the few \% range.

\section{Decay constants}

We define the meson (axial-vector) decay constant in finite volume as  
\begin{equation}
\label{defFpi}
\left< 0 |A_\mu^M | M(p) \right> = i\sqrt{2}F_M p_\mu + i\sqrt{2}F^V_{M\mu}\,,
\end{equation}
where $M(p)$ is a meson and 
$A_\mu=\bar q \gamma_\mu\gamma_5({\lambda^M}/{\sqrt2}) q$ 
is the axial current. The extra term is needed
since the matrix element in finite volume is no longer proportional to $p_\mu$.
The first term in (\ref{defFpi})
can be identified by looking at the time component of
the current. The second term has non-zero components only in the
spatial directions and vanishes in infinite volume.

For the flavour charged mesons, the charge in the axial current and
the meson is necessarily the same. In the isospin limit
the same is true for the $\pi^0$ and the $\eta$. However the twisted boundary
conditions do break isospin and thus the $\pi^0$ also couples to
the octet current and the $\eta$ to the triplet current.
At NLO this coupling comes from two effects, the mixing between the
isospin triplet $\pi$ and the octet $\eta$ as well as the direct
transition to the other current. A derivation can be found in Sect.~2.2
of \cite{ABT2}.

We also consider decay through a pseudo-scalar current. We define this decay
constant as
\begin{equation}
  \left< 0|P^M|M(p)\right> = \frac{G_M}{\sqrt{2}}
\end{equation}
where $P=\bar q i\gamma_5(\lambda^M/\sqrt2)q$ is the pseudo-scalar current
corresponding to the meson $M$. A similar comment to above about
$\pi^0$ and $\eta$ applies.

These two matrix elements satisfy the Ward identity
\begin{equation}
\partial^\mu \left< 0 |A_\mu^M | M(p) \right>
= (m_q+m_{q'}) \left< 0|P^M|M(p)\right>\,,
\end{equation}
valid for flavour charged mesons of composition $\bar q q'$. 
This leads to 
\begin{equation}
  p^2 F_M+p^\mu F^V_{M\mu} = \frac{1}{2}(m_q+m_{q'})G_M\,.
  \label{eq:wardFG}
\end{equation}
We have checked that our expressions for the charged mesons agree with this.
An important part in this agreement is the use
of the correct momentum-dependent mass of the meson.
For the neutral mesons a somewhat more complicated relation is needed
since they are sums of terms with different quark masses.

The analytical results for the finite volume effects
on the axial-vector decay constants are given below 
in terms of the integrals defined in App.~\ref{app:Integrals}.
For the $\pi^0$ and $\eta$ we listed the matrix-elements
with $A_\mu^3$ and $A_\mu^8$ separately, indicating which decay is which with
an extra subscript. The isospin breaking decay vanishes if the up and down
quarks have the same twist angles.

Again we agree with the infinite volume results of \cite{GL2}.
The finite volume corrections for the axial current decay constants
for the flavour charged mesons are
\begin{align}
\label{axialresultcharged}
\Delta^V\! F_{\pi^\pm} &= \frac{1}{F_0}\left(
 \frac{1}{2} A^V(m_{\pi^+}^2) + \frac{1}{2} A^V(m_{\pi^0}^2) 
+ \frac{1}{4}A^V(m_{K^+}^2) + \frac{1}{4}A^V(m_{K^0}^2)\right)\,,
\nonumber\\
F^V_{\pi^\pm\mu} &=
  \pm\frac{1}{F_0}\left[ 2 A^V_\mu(m_{\pi^+}^2) +  A^V_\mu(m_{K^+}^2) -  A^V_\mu(m_{K^0}^2)\right]\,,
\nonumber\\
\Delta^V\! F_{K^\pm} &= \frac{1}{F_0}\left(
    \frac{1}{4}A^V(m_{\pi^+}^2) + \frac{1}{8} A^V(m_{\pi^0}^2)
    + \frac{1}{2}A^V(m_{K^+}^2) + \frac{1}{4}A^V(m_{K^0}^2)
    + \frac{3}{8} A^V(m_{\eta}^2)\right)\,,
\nonumber\\
F^V_{K^\pm \mu} &=\pm\frac{1}{F_0}\left[A^V_\mu(m_{\pi^+}^2) + 2 A^V_\mu(m_{K^+}^2) +  A^V_\mu(m_{K^0}^2)\right]\,,
\nonumber\\
\Delta^V\! F_{K^0(\bar{K}^0)}&= \frac{1}{F_0}
\left(\frac{1}{4}A^V(m_{\pi^+}^2) + \frac{1}{8}A^V(m_{\pi^0}^2)
   + \frac{1}{4} A^V(m_{K^+}^2) + \frac{1}{2}A^V(m_{K^0}^2)
   +  \frac{3}{8} A^V(m_{\eta}^2)\right)\,,
\nonumber\\
F^V_{K^0(\overline K^0)\mu} &=
    +(-) \frac{1}{F_0}\left[-  A^V_\mu(m_{\pi^+}^2) +  A^V_\mu(m_{K^+}^2) + 2 A^V_\mu(m_{K^0}^2)\right]\,.
\end{align}
They agree with the untwisted finite volume results of \cite{Becirevic}. 
The relation to the results given in \cite{twisted} is discussed in
Sect.~\ref{sec:comparison}.
The flavour neutral
expressions include the effects of mixing.
\begin{align}
\label{axialresultneutral}
F^V_{\pi^0 3\mu} &= F^V_{\pi^0 8\mu} = F^V_{\eta 3\mu}=F^V_{\eta 8\mu} = 0\,,
\nonumber\\
\Delta^V\! F_{\pi^0 3} &= \frac{1}{F_0}( A^V(m_{\pi^+}^2)
   + \frac{1}{4}A^V(m_{K^+}^2) + \frac{1}{4}A^V(m_{K^0}^2))\,,
\nonumber\\
\Delta^V\! F_{\pi^0 8} &=
   \frac{3m_\eta^2-m_\pi^2}{2\sqrt{3}F_0(m_\eta^2-m_\pi^2)}( A^V(m_{K^+}^2) - A^V(m_{K^0}^2))\,,
\nonumber\\
\Delta^V\! F_{\eta8} &= \frac{3}{4F_0}( A^V(m_{K^+}^2) + A^V(m_{K^0}^2))\,,
\nonumber\\
\Delta^V\! F_{\eta3} &= \frac{-m_\pi^2}{\sqrt{3}F_0(m_\eta^2-m_\pi^2)}( A^V(m_{K^+}^2) - A^V(m_{K^0}^2)).
\end{align}
to simplify the expressions.

The masses and $F_0$ in these expressions can be chosen in different ways
as discussed earlier for the masses.

The lowest order value for the pseudo-scalar decay constants is
 $G_0 = 2 F_0 B_0$.
We are not aware of published results for the
NLO corrections at infinite volume, we thus quote those for completeness
and add a superscript
$(4)$ to indicate the NLO infinite volume correction. 
Note that isospin is valid at infinite volume such that the
mixed ones vanish and there is only an expression for the $\pi$, $K$ and
$\eta_8$ case.
\begin{align}
G^{(4)}_\pi &=
\frac{G_0}{F_0^2}\left(4 K_{46}
    +4m_\pi^2(4L_8^r-L_5^r)
    +\frac{1}{2}\overline{A}(m_\pi^2)
    +\frac{1}{2}\overline{A}(m_K^2)
    +\frac{1}{6}\overline{A}(m_\eta^2)\right)\,,
\nonumber\\
G^{(4)}_K &=
\frac{G_0}{F_0^2}\left(4 K_{46}
    +4m_K^2(4L_8^r-L_5^r)
    +\frac{3}{8}\overline{A}(m_\pi^2)
    +\frac{3}{4}\overline{A}(m_K^2)
    +\frac{1}{24}\overline{A}(m_\eta^2)\right)\,,
\nonumber\\
G^{(4)}_{\eta8} &=
\frac{G_0}{F_0^2}\left(4 K_{46}
    +4m_\eta^2(4L_8^r-L_5^r)
    +\frac{1}{2}\overline{A}(m_\pi^2)
    +\frac{1}{6}\overline{A}(m_K^2)
    +\frac{1}{2}\overline{A}(m_\eta^2)\right)\,,
\nonumber\\
K_{46} &=(2m_K^2+m_\pi^2)(4L_6^r-L_4^r)\,. 
\end{align}
The integral
is
\begin{align}
\overline A(m^2) &= -\frac{m^2}{16\pi^2}\log\frac{m^2}{\mu^2}\,.
\end{align}

The finite volume effects for the pseudo-scalar decay constants 
for the flavour charged mesons are
\begin{align}
\Delta^V\! G_{\pi^\pm}^V &= \frac{G_0}{F_0^2}
  \left(\frac{1}{2}A^V(m_{\pi^+}^2) + \frac{1}{4}A^V(m_{K^+}^2) 
 + \frac{1}{4}A^V(m_{K^0}^2) + \frac{1}{6}A^V(m_{\eta}^2)\right)\,,
\nonumber\\
\Delta^V\! G_{K^\pm} &= \frac{G_0}{F_0^2}
\left( \frac{1}{4}A^V(m_{\pi^+}^2) + \frac{1}{8}A^V(m_{\pi^0}^2)
 +\frac{1}{2} A^V(m_{K^+}^2) + \frac{1}{4}A^V(m_{K^0}^2)
 + \frac{1}{24}A^V(m_{\eta}^2)\right)\,,
\nonumber\\
\Delta^V\! G_{K^0(\overline K^0)} &= \frac{G_0}{F_0^2}
\left( \frac{1}{4}A^V(m_{\pi^+}^2) + \frac{1}{8}A^V(m_{\pi^0}^2)
 +\frac{1}{4} A^V(m_{K^+}^2) + \frac{1}{2}A^V(m_{K^0}^2)
 + \frac{1}{24}A^V(m_{\eta}^2)\right)\,.
\end{align}
For the flavour neutral cases we need to take into account mixing and obtain
\begin{align}
 \Delta^V\! G_{\pi^0 3} &= \frac{G_0}{F_0^2} \left(\frac{1}{2} A^V(m_{\pi^0}^2)
 + \frac{1}{4}A^V(m_{K^+}^2) + \frac{1}{4}A^V(m_{K^0}^2)
 + \frac{1}{6}A^V(m_{\eta}^2)\right)\,,
\nonumber\\
 \Delta^V\! G_{\pi^0 8} &=\frac{G_0}{F_0^2}
\frac{m_\eta^2+m_\pi^2}{2\sqrt{3}(m_\eta^2-m_\pi^2)}
\left( A^V(m_{K^+}^2) - A^V(m_{K^0}^2)\right)\,,
\nonumber\\
\Delta^V\! G_{\eta 8} &= \frac{G_0}{F_0^2} 
\left( \frac{1}{3}A^V(m_{\pi^+}^2) + \frac{1}{6}A^V(m_{\pi^0}^2)
 + \frac{1}{12}A^V(m_{K^+}^2) + \frac{1}{12}A^V(m_{K^0}^2)
 +\frac{1}{2} A^V(m_{\eta}^2)\right)\,,
\nonumber\\
 \Delta^V\! G_{\eta 3} &=\frac{G_0}{F_0^2}
\frac{-m_\eta^2}{\sqrt{3}(m_\eta^2-m_\pi^2)}
\left( A^V(m_{K^+}^2) - A^V(m_{K^0}^2)\right)\,.
\end{align}
At this order $G_{\pi^08}$ and $G_{\eta3}$
only arise from $\pi^0$-$\eta$ mixing.

We present now some numerics for the same inputs as used for the
masses given in (\ref{input1}) and (\ref{input2}).

\begin{figure}
\begin{center}
\includegraphics[width=\figWidth\textwidth]{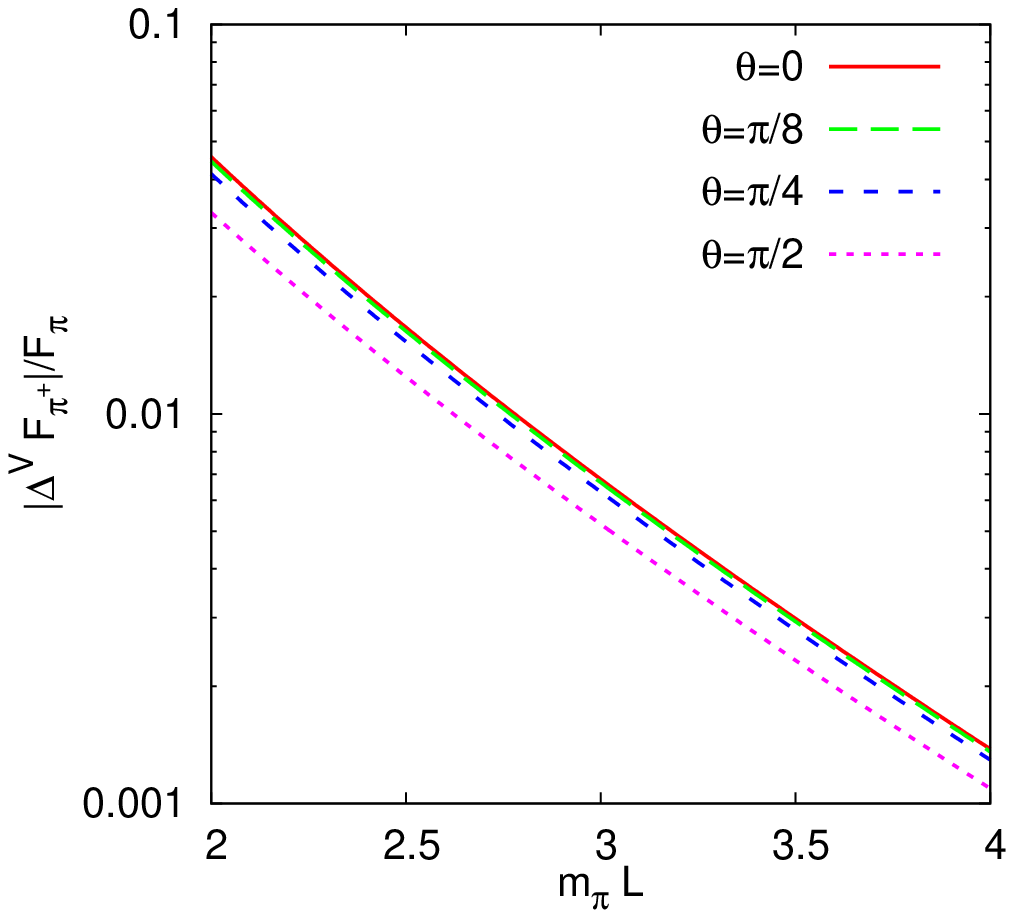}
\hspace*{0.05\textwidth}
\includegraphics[width=\figWidth\textwidth]{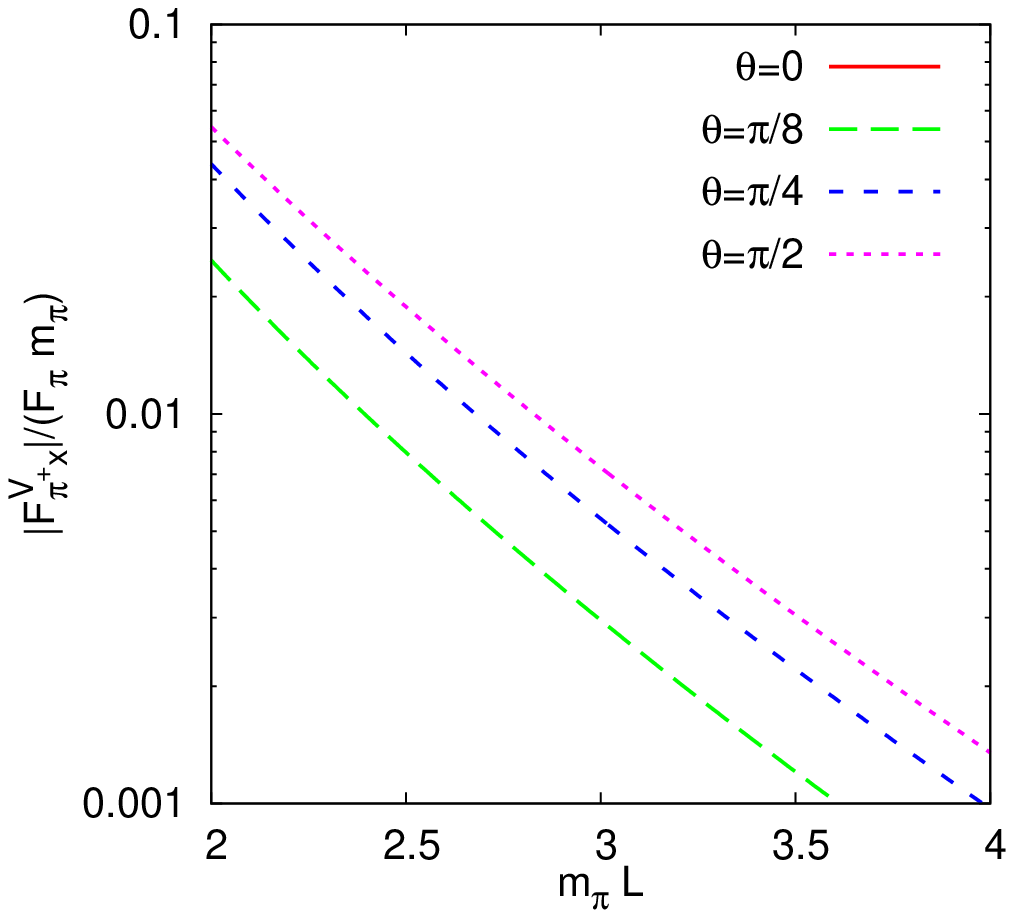}\\
\includegraphics[width=\figWidth\textwidth]{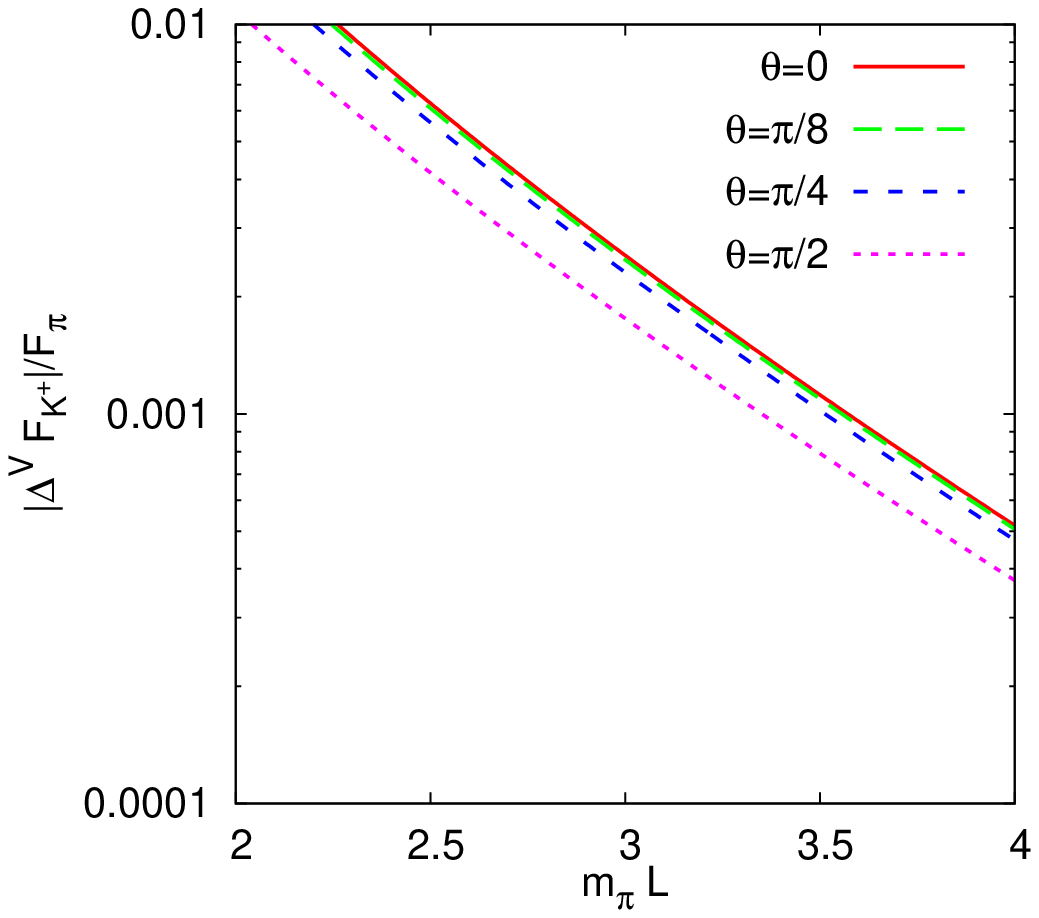}
\hspace*{0.05\textwidth}
\includegraphics[width=\figWidth\textwidth]{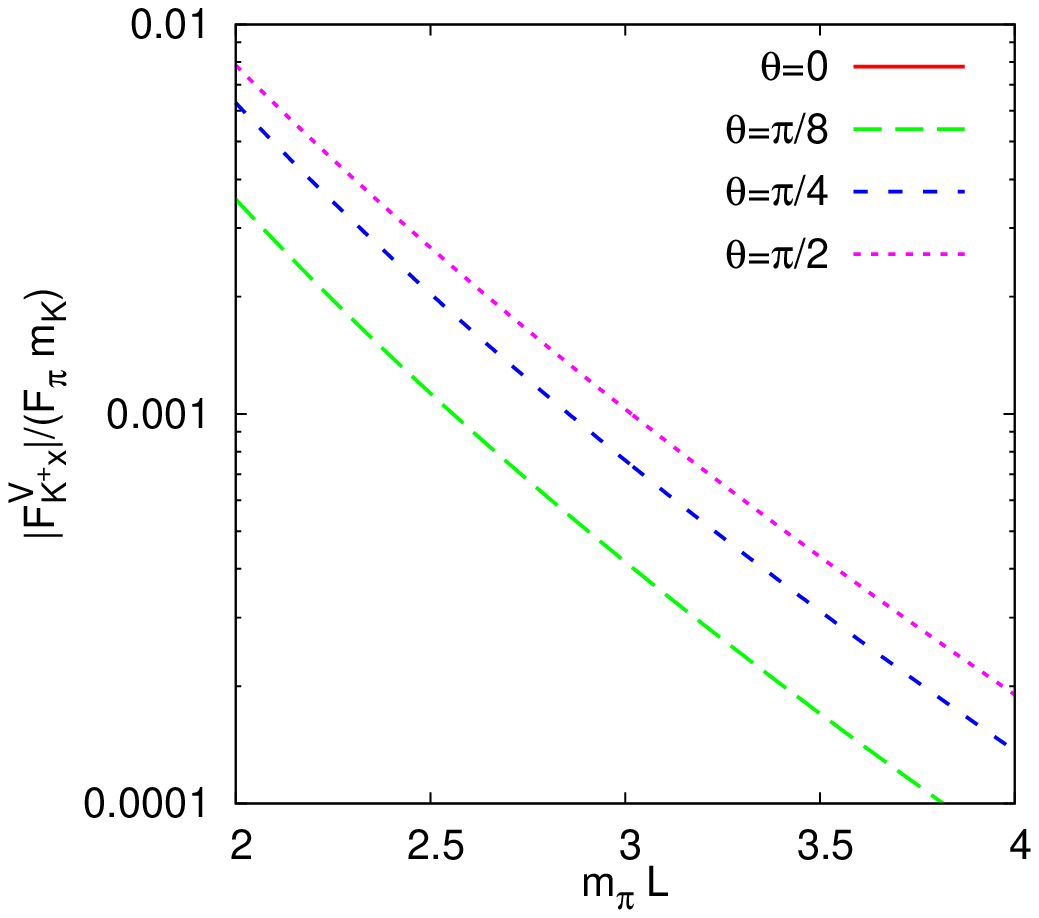}
\end{center}
\caption{\label{figDecay}
Relative finite volume correction for the two terms
in the decay constant matrix element (\ref{defFpi}).
On the left hand side we have plotted $\Delta^V F_M/F_\pi$
and on the right hand side $F^V_{Mx}/(F_\pi m_M)$, i.e. the $x$-component
compared to the size of the zero-component. For the input chosen the
$x$-component is the only non-zero one for the second term in (\ref{defFpi}).
The top row is $M=\pi^+$
and the bottom row for $M=K^+$. Input values as in (\ref{input1})
and (\ref{input2}).}
\end{figure}

In Fig.~\ref{figDecay} we show the size of the finite volume corrections
to the charged meson decay constants with both terms in (\ref{defFpi})
shown separately. We use the same input parameters as for the masses
of (\ref{input1}) and (\ref{input2}).
The first term in (\ref{defFpi})
is shown in the left plots normalized to $F_\pi$ for
the charged pion and kaon. The right plots shows the $x$-component of the
second term in (\ref{defFpi}), which is the only non-zero component for our
choice of input. It vanishes identically for $\theta=0$. We have normalized
here to the value of $F_\pi m_K$ which is roughly
the value of the $t$-component in infinite volume.
Note that the finite volume corrections can be sizable and the second term
is not always negligible.

\section{Electromagnetic form-factor}
\label{sec:formfactor}

The electromagnetic form-factor in infinite volume is defined as
\begin{equation}
  \left<p^\prime|j_\mu^{em}|p\right> =  F(q^2)(p+p^\prime)_\mu
  \label{eq:EMinfinite}
\end{equation}
where $q = p-p^\prime$ and $j^\mu$ is the electromagnetic current
for the light quark flavours
\begin{equation}
  j_\mu^{em} = \frac{2}{3}\bar{u}\gamma_\mu u - \frac{1}{3}(\bar{d}\gamma_\mu d + \bar{s}\gamma_\mu s).
\end{equation}
The electromagnetic form-factor in twisted lattice QCD is not the same
as in infinite volume or finite volume with periodic conditions. 
Instead it has the more general form
\begin{align}
\label{deffh}
  \left<M^\prime(p^\prime)|j_\mu^I|M(p)\right> &= f_{IMM^\prime  \mu}
\nonumber\\
  &= f_{IMM^\prime  +} (p_\mu+p_\mu^\prime) 
    + f_{IMM^\prime  -} q_\mu
    +  h_{IMM^\prime  \mu}\,.
\end{align}
In addition to the electromagnetic current we will use
\begin{align}
j^q_\mu &= \bar q\gamma_\mu q, & j^{\pi^+}_\mu &= \bar d\gamma_\mu u\,.
\end{align}
We will also suppress the $M^\prime$ in the subscripts when initial and
final meson are the same and sometimes the $IMM^\prime $.
In the infinite volume limit the functions $f_-$ and $h$ must go to zero and
$f_+$ must go to $F(q^2)$ so that Eq. (\ref{eq:EMinfinite}) is recovered.
We only work with currents where the quark and anti-quark have the same mass.
The result in infinite volume can be found in \cite{GL3}.
Results at finite volume with periodic boundary conditions are in
\cite{Bunton:2006va,Ghorbani}.

The main reason for using twisted boundary conditions is to extract physical
quantities for small momenta. In the case of the electromagnetic form-factor
the twist does not help when applied to correlators such as
\begin{equation}
  \left<\pi^+(p^\prime)|j_\mu^q|\pi^+(p)\right>
\end{equation}
since the same twist is applied to the incoming and outgoing particles we
get $p_i-p^\prime_i = 2\pi n_i/L$. However, as was pointed out in
\cite{UKQCD1}, it is possible to extract information using isospin symmetry.
To analyze this more carefully requires calculations in
partially quenched \chpt\ and this will be the topic of forthcoming work. 
Here we are satisfied with noting that in the isospin limit
with $m_u=m_d$ and $\theta_u=\theta_d$ we have the relation
(in our sign conventions)
\begin{equation}
\label{eq:isorelation}
\left<\pi^+(p^\prime)|\bar{u}\gamma_\mu u |\pi^+(p)\right> =
 -\left<\pi^+(p^\prime)|\bar{d}\gamma_\mu d |\pi^+(p)\right> = 
-\frac{1}{\sqrt2}\left<\pi^0(p^\prime)|\bar{d}\gamma_\mu u |\pi^+(p)\right>.
\end{equation}
The relation
(\ref{eq:isorelation}) can in principle be used to evaluate the main part,
excluding $\bar s\gamma_\mu s$,
of the electromagnetic form-factor
of the pion for arbitrary momenta.
The currents $\bar{d}\gamma_\mu u$ is referred to as $\bar du$ in the
equations below.
In practice $\pi^0$ gives rise to
difficulties on the lattice, and the twisted boundary conditions explicitly
break isospin. The corrections due to the latter are
one of the goals of this work.

\subsection{Analytic expressions}
\label{sec:analyticEM}

The split in $f_+,f_-$ and $h$ in (\ref{deffh}) is not unique. The functions
can depend on all components of the momenta and twist-vectors.
However, we stick to the splitting among $f_+,f_-$ and $h$ which 
naturally emerges from the one-loop calculation.
The integrals appearing are defined in App.~\ref{app:Integrals}.

The results for $f_+^V$ are most easily given in terms of the finite volume
generalization of the function $\mathcal{H}$ in \cite{GL3,BT}.
\begin{align}
H^V(m_1^2,m_2^2,q) &=
\frac{1}{4}A^V(m_1^2)+\frac{1}{4}A^V(m_2^2)-B_{22}^V(m_1^2,m_2^2,q)
\end{align}
The effects of $\pi^0$-$\eta$ mixing appear earliest at NNLO for the
form-factors listed here.
The form-factors $f_+$ we consider are:
\begin{align}
\label{resultf+}
\Delta^V\! f_{em \pi^\pm +} &=\frac{\pm1}{F_0^2}\left( 2 H^V(m_{\pi^+}^2,m_{\pi^-}^2,q)
                           + H^V(m_{K^+}^2,m_{K^-}^2,q)\right)\,,
\nonumber\\
\Delta^V\! f_{em K\pm +} &=\frac{\pm1}{F_0^2}\left(  H^V(m_{\pi^+}^2,m_{\pi^-}^2,q)
                           +2 H^V(m_{K^+}^2,m_{K^-}^2,q)\right)\,,
\nonumber\\
\Delta^V\! f_{em K^0(\overline K^0) +} &=\frac{\pm1}{F_0^2}\left( - H^V(m_{\pi^+}^2,m_{\pi^-}^2,q)
                           + H^V(m_{K^+}^2,m_{K^-}^2,q)\right)\,,
\nonumber\\
\Delta^V\! f_{em \pi^0+} &= 0\,,
\nonumber\\
\Delta^V\! f_{\bar du \pi^+\pi^0 +} &=\frac{-\sqrt{2}}{F_0^2}\left( 2 H^V(m_{\pi^+}^2,m_{\pi^0}^2,q)
                           + H^V(m_{K^+}^2,m_{\overline K^0}^2,q)\right)\,.
\end{align}
The $f_-$ form-factors for the same cases are:
\begin{align}
\label{resultf-}
\Delta^V\! f_{em \pi^+(\pi^-) -} &=
\frac{p^{\prime\nu}(-p^\nu)}{F_0^2}\left(2 B^V_{2\nu}(m_{\pi^+}^2,m_{\pi^-}^2,q)
                           + B^V_{2\nu}(m_{K^+}^2,m_{K^-}^2,q)\right)\,,
\nonumber\\
\Delta^V\! f_{em K^+(K^-) -} &=
\frac{p^{\prime\nu}(-p^\nu)}{F_0^2}\left( B^V_{2\nu}(m_{\pi^+}^2,m_{\pi^-}^2,q)
                           + 2B^V_{2\nu}(m_{K^+}^2,m_{K^-}^2,q)\right)\,,
\nonumber\\
\Delta^V\! f_{em K^0(\overline K^0) -} &=
\frac{1}{F_0^2}\left(-(p^{\nu}(-p^{\prime\nu})) B^V_{2\nu}(m_{\pi^+}^2,m_{\pi^-}^2,q)
                           + p^{\prime\nu}(-p^\nu)B^V_{2\nu}(m_{K^+}^2,m_{K^-}^2,q)\right)\,,
\nonumber\\
\Delta^V\! f_{em \pi^0 -} &=\frac{1}{F_0^2}
\bigg(m_\pi^2\left(B^V(m_{\pi^+}^2,m_{\pi^-}^2,q)-2B_1^V(m_{\pi^+}^2,m_{\pi^-}^2,q)\right)
\nonumber\\*
&-q^\nu\left(2B^V_{2\nu}(m_{\pi^+}^2,m_{\pi^-}^2,q)
                +\frac{1}{2}B^V_{2\nu}(m_{K^+}^2,m_{K^-}^2,q)\right)\bigg)\,,
\nonumber\\
\Delta^V\! f_{\bar du\pi^+ \pi^0 -} &=\frac{\sqrt{2}}{F_0^2}
\bigg(m_\pi^2\left(B^V(m_{\pi^+}^2,m_{\pi^0}^2,q)-2B_1^V(m_{\pi^+}^2,m_{\pi^0}^2,q)\right)
\nonumber\\*
&-\left(2p^\nu B^V_{2\nu}(m_{\pi^+}^2,m_{\pi^-}^2,q)
                +\frac{1}{2}(p+p^\prime)^\nu B^V_{2\nu}(m_{K^+}^2,m_{\overline K^0}^2,q)\right)\bigg)\,,
\end{align}  
Finally, the $h_\mu$ at finite volume are
\begin{align}
\label{resulth}
\Delta^V\! h_{em \pi^\pm\mu} &=
\frac{1}{F_0^2}\bigg( 2A^V_\mu(m_{\pi^+}^2)+A^V_\mu(m_{K^+}^2)-A^V_\mu(m_{K^0}^2)
\nonumber\\*
&
+q^2B^V_{2\mu}(m_{\pi^+}^2,m_{\pi^-}^2,q)
+\frac{q^2}{2}B^V_{2\mu}(m_{K^+}^2,m_{K^-}^2,q)
\nonumber\\*
&
\mp(p+p^\prime)^\nu\left(2B^V_{23\mu\nu}(m_{\pi^+}^2,m_{\pi^-}^2,q)
+B^V_{23\mu\nu}(m_{K^+}^2,m_{K^-}^2,q)\right)\bigg)\,,
\nonumber\\
\Delta^V\! h_{em K^\pm\mu} &=
\frac{1}{F_0^2}\bigg( A^V_\mu(m_{\pi^+}^2)+2A^V_\mu(m_{K^+}^2)+A^V_\mu(m_{K^0}^2)
\nonumber\\*
&+\frac{q^2}{2}B^V_{2\mu}(m_{\pi^+}^2,m_{\pi^-}^2,q)
+q^2B^V_{2\mu}(m_{K^+}^2,m_{K^-}^2,q)
\nonumber\\*
&\mp(p+p^\prime)^\nu\left(B^V_{23\mu\nu}(m_{\pi^+}^2,m_{\pi^-}^2,q)
+2B^V_{23\mu\nu}(m_{K^+}^2,m_{K^-}^2,q)\right)\bigg)\,,
\nonumber\\
\Delta^V\! h_{em K^0(\overline K^0)\mu} &=
\frac{1}{F_0^2}\bigg(
 \frac{q^2}{2}B^V_{2\mu}(m_{\pi^+}^2,m_{\pi^-}^2,q)
+\frac{q^2}{2}B^V_{2\mu}(m_{K^+}^2,m_{K^-}^2,q)
\nonumber\\*
&
+(-)(p+p^\prime)^\nu\left(B^V_{23\mu\nu}(m_{\pi^+}^2,m_{\pi^-}^2,q)
-B^V_{23\mu\nu}(m_{K^+}^2,m_{K^-}^2,q)\right)\bigg)\,,
\nonumber\\
\Delta^V\! h_{em \pi^0\mu} &=
\frac{1}{F_0^2}\bigg(
2(q^2-m_\pi^2)B^V_{2\mu}(m_{\pi^+}^2,m_{\pi^-}^2,q)
+\frac{q^2}{2}B^V_{2\mu}(m_{K^+}^2,m_{K^-}^2,q)\bigg)\,,
\nonumber\\
\Delta^V\! h_{\bar du\pi^+\pi^0\mu} &=
\frac{\sqrt{2}}{F_0^2}\bigg(
-A^V_\mu(m_{\pi^+}^2)-\frac{1}{2}A^V_\mu(m_{K^+}^2)+\frac{1}{2}A^V_\mu(m_{K^0}^2)
\nonumber\\*
&+(q^2-2m_\pi^2)B^V_{2\mu}(m_{\pi^+}^2,m_{\pi^0}^2,q)\bigg)\,,
\nonumber\\*
&+(p+p^\prime)^\nu\left(2B^V_{23\mu\nu}(m_{\pi^+}^2,m_{\pi^0}^2,q)
+B^V_{23\mu\nu}(m_{K^+}^2,m_{\overline K^0}^2,q)\right)\bigg)\,.
\end{align}
We used in these formulas that the $\pi^0$ and $\eta$ have no twist and
that particle and anti-particle have opposite twists.
Both $f_-$ and $h$ vanish in infinite volume.

\subsection{Ward identities}

All the form-factors we discuss have the same mass for the quark and anti-quark
in the vector current.
As a consequence they obey, even at finite volume, the Ward identity
\begin{align}
q^\mu f_{I MM^\prime \mu}
&= (p^2-p^{\prime'2}) f_{I MM^\prime +} + q^2 f_{I MM^\prime -} +q^\mu h_{I MM^\prime \mu} =0\,.
\end{align}
We have used this as a check on our results.
This standard check requires a bit of caution when using twisted boundary
conditions. The issue is that masses are momentum dependent
when twist is applied, see Sect.~\ref{sec:mass}. When performing a
one loop calculation part of the mass correction
is different for ingoing and outgoing meson, this means that
$p^2-p^{\prime2}\ne0$ even when the incoming and outgoing particle
are the same.
Comparing equations for the mass corrections, we see that these cancel
the parts coming from $A^V_\mu$ in $h_{IMM^\prime \mu}$. The remainder cancels
between $q^2f_{IMM^\prime -}$ and  $q^\mu h_{IMM^\prime \mu}$ when using
the identities in App.~\ref{app:relations}. 

\subsection{Numerical results}

Let us first remind here why twisting is useful for form-factors with
the example of the pion form-factor and a lattice size of $m_\pi L=2$.
The smallest spatial momentum that can be produced is $2\pi/L = \pi m_\pi$
and the corresponding $q^2$ is $q^2_{min} = -0.089~\mathrm{GeV}^2
= -(0.3~\mathrm{GeV})^2$.
Twisting allows for $q^2$ continuously varying from zero.

In this section we concentrate on the quantity
\begin{equation}
\label{deffV}
f_\mu = \left(1+f^{\infty}_++\Delta^V f_+ \right)(p+p^\prime)_\mu
   +\Delta^V f_- q_\mu + \Delta^V h_\mu =
 -\frac{1}{\sqrt{2}} f_{\bar du\pi^+\pi^0\mu}\,.
\end{equation}
This is the form-factor corresponding to the right hand side of
(\ref{eq:isorelation}) normalized to 1 at $q^2=0$ in infinite volume.
The finite volume parts are what is needed to obtain the pion
electromagnetic form-factor, neglecting the $s$-quark contribution,
at infinite volume. We have separated the lowest order value of 1,
the infinite volume and finite volume correction to $f_+$ as well as the
$f_-$ and $h_\mu$ parts defined earlier.

\begin{figure}
\begin{center}
\includegraphics[width=\figWidth\textwidth]{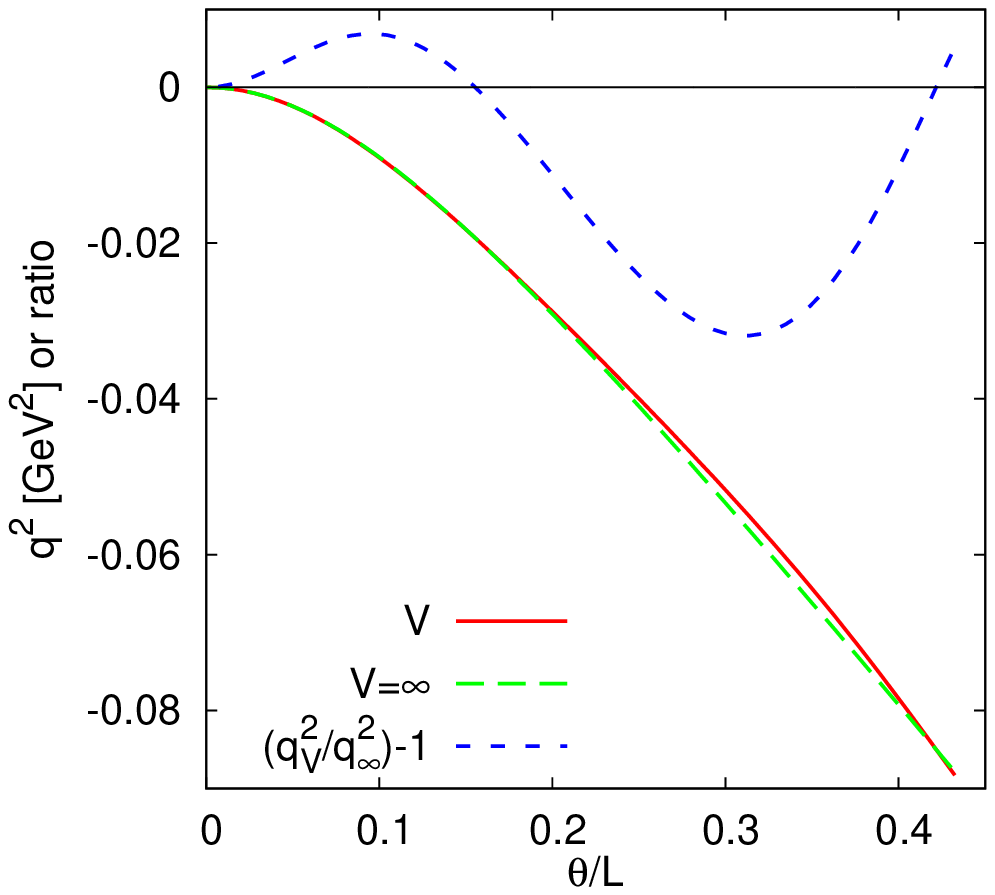}
\hspace{0.05\textwidth}
\includegraphics[width=\figWidth\textwidth]{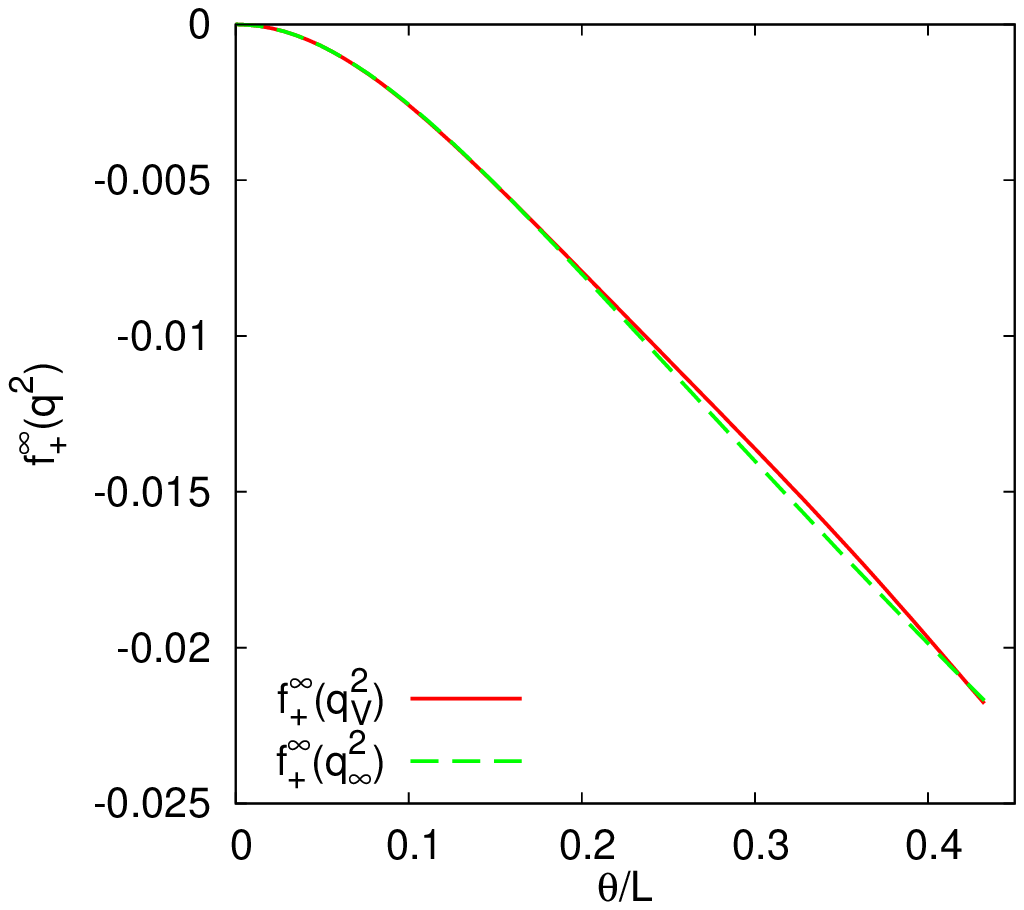}
\end{center}
\caption{\label{figqsq} Left: The dependence of $q^2$ at a fixed
$\vec q=(\theta/L,0,0)$ for the finite volume with $m_\pi L=2$ and infinite
volume as well as the difference ratio from one. 
The curves end at $\theta=2\pi$.
Right: The effect of this change in $q^2$ on the infinite volume corrections
of $f_+^V(q^2)$ with $L_9^r=0$.}
\end{figure}

Again we look at the case with $\vec\theta_u=(\theta,0,0)$.
This means that the incoming $\pi^+$ four-momentum $p$, the outgoing
$\pi^0$ momentum $p^\prime$ and $q^2$ are
\begin{align}
p &= \left(\sqrt{m_{\pi^+}^{V2}+(\theta/L)^2},\theta/L,0,0\right)\,,
\nonumber\\
p^\prime &= \left(m_{\pi^0}^{V2},0,0,0\right)\,,
\nonumber\\
q^2 &= m_{\pi^+}^{V2}+m_{\pi^0}^{V2}-2m_{\pi^0}^V\sqrt{m_{\pi^+}^{V2}+(\theta/L)^2}\,.
\end{align}
Note that the masses at finite volume that come in here,
not the infinite volume ones. We have indicated this with the superscript
$V$ in the masses.
To plot the corrections we use $m_M^{V2} = m_M^2+\Delta^V m_M^2$
in the numerics with $\Delta^V m_M^2$ given in (\ref{resultmasses}).
The size of this effect is shown in the left plot of Fig.~\ref{figqsq}. 
We plot the
value of $q^2$ at finite and infinite volume and the 
deviation of the ratio from 1 as a function
of $\theta/L$. The endpoint of the curve is for $\theta=2\pi$.
The right plot in Fig.~\ref{figqsq} shows the effect on the form-factor
of this change in $q^2$. We plotted there the one-loop contribution
at infinite volume
to the pion electromagnetic form-factor, $f_+^\infty(q^2)$,
as a function of the two different $q^2$
discussed here. The extra input values
used are $L_9^r=0$ and $\mu=0.77~\mathrm{GeV}$.
The total effect of this correction is rather small.

\begin{figure}
\begin{center}
\includegraphics[width=\figWidth\textwidth]{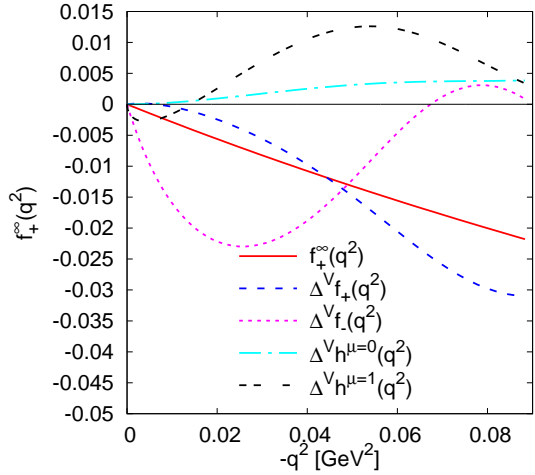}
\end{center}
\caption{\label{figfvparts} The various parts of the form-factor
defined in (\ref{deffV}). See text for a more detailed explanation.}
\end{figure}

In the remainder we will use the $q^2$ as calculated with the finite
volume masses. In Fig.~\ref{figfvparts} we plot the different parts
of the form-factor as defined in (\ref{deffV}).
Plotted are the infinite volume one-loop part of $f_+^\infty$,
the finite volume corrections $\Delta^V f_+$, $\Delta^V f_-$
and the two non-zero components of $\Delta^V h^\mu$. As one can see,
the finite volume corrections are not small and the parts due to the extra
form-factors can definitely not be neglected. The units are GeV for the
two components of $\Delta^V h^\mu$.

\begin{figure}
\begin{center}
\includegraphics[width=\figWidth\textwidth]{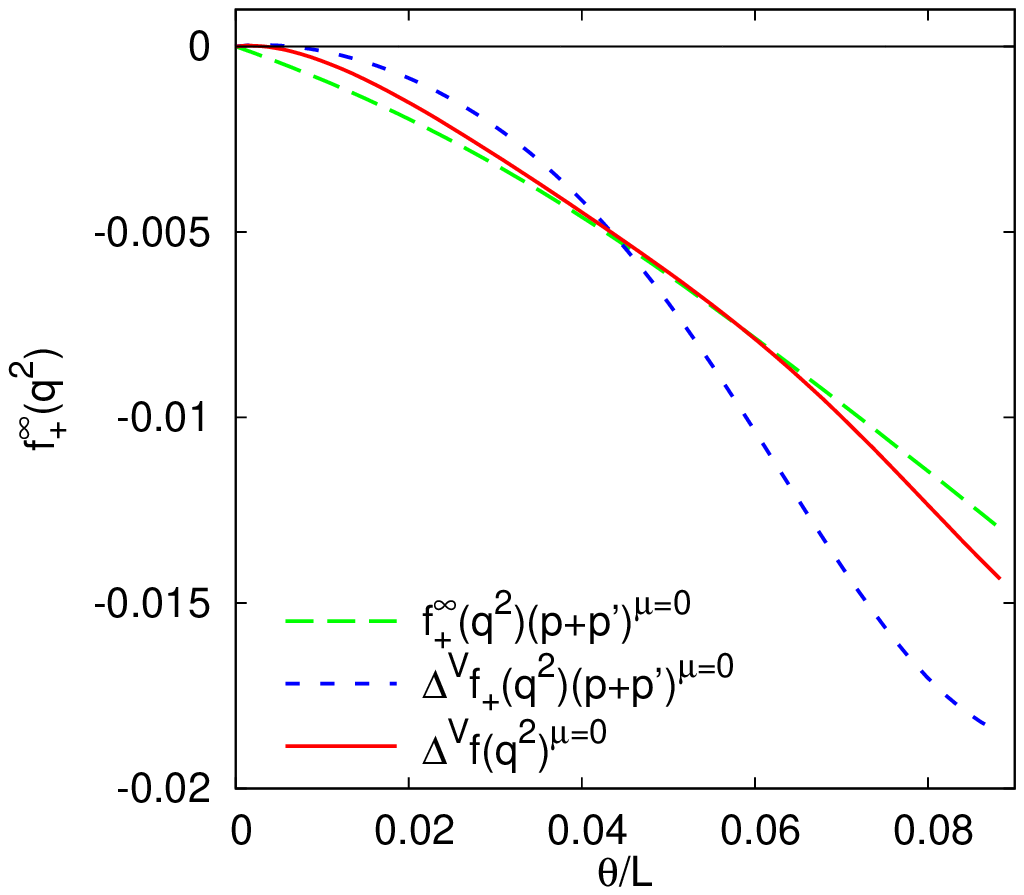}
\hspace{0.05\textwidth}
\includegraphics[width=\figWidth\textwidth]{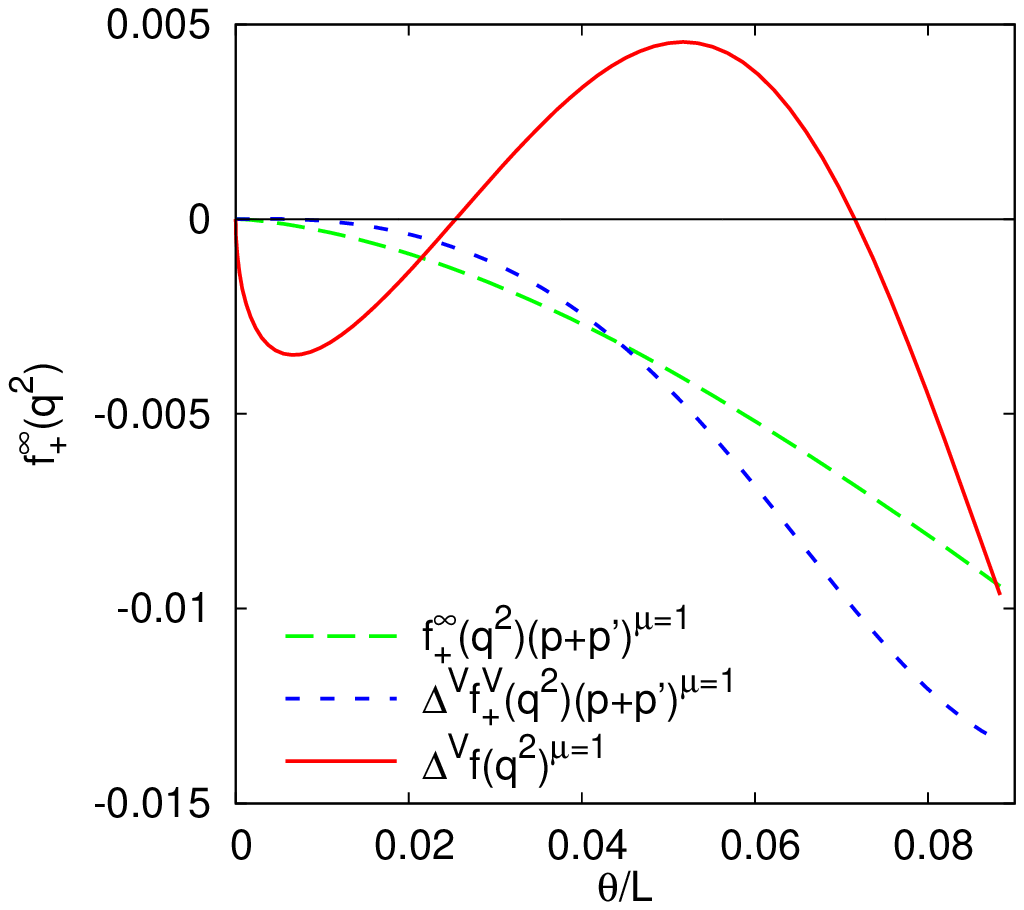}
\end{center}
\caption{\label{figfvcomp} Left: $\mu=0$ Right: $\mu=1$.
Plotted are those due to the one-loop infinite volume correction,
$f_+^\infty(q^2)$, the finite volume
correction to $f_+$, $\Delta^V f_+$,
and the full finite volume correction, 
$\Delta^V f^\mu = \Delta^V f_+(p+p^\prime)^\mu+ \Delta^V f_-q^\mu + \Delta^V h^\mu$.}
\end{figure}

The more relevant quantities for comparison are the components with $\mu=0$
and $\mu=1$. We have plotted the form-factor as defined with upper index $\mu$.
The left plot in Fig.~\ref{figfvcomp} shows $\mu=0$ and the right plot
$\mu=1$. Units are in GeV. The finite volume correction is of a size similar
to the infinite volume pure one-loop contribution and the correction due to
the extra terms at finite volume and twist are not negligible.

\section{Comparison with earlier work}
\label{sec:comparison}

The one and two-point Green functions of vector currents are discussed
in Sect.~\ref{sec:vector}. These issues were discussed in a more lattice
oriented way in \cite{Aubin}. Here we have provided the \chpt\ expressions
for them.

For the masses the comparison with earlier work is more subtle.
In this work, we have consistently used the formulation with
non-zero twist angle and no induced background field. This implies that
the allowed meson momenta are of the form
$\vec p_{BR} = (2\pi\vec n+\vec\theta)/L$, with $\vec n$ a three-vector with
integer components and $\vec\theta$ the twist vector for the field
corresponding to the meson.
As mentioned in Sect.~\ref{sec:chpt} we define asymptotic states
as those where there is at fixed $\vec p$ a pole at a value, $E_0$, of the
energy. The LSZ theorem can then be used for these single particle states to
obtain matrix elements by taking the limit $E\to E_0$ allowing for
the usual method with wave function renormalization and possibly mixing
of external states to take into account external leg corrections.
Our definition of the mass used is
\begin{equation}
\label{defmassBR}
m_{BR}^2 = E^2_0-\vec p_{BR}^2\,.
\end{equation}
The mass can depend on all components of $\vec p$ since there is no rotation
invariance and even cubic invariance\footnote{We assume here that the $t$
direction is infinite.} is no longer present. We have used the expression
``momentum-dependent mass'' in the text to indicate this dependence. The
relation between $E$ and $\vec p$ for states is called
dispersion relation in some other references, see e.g. \cite{JT}.

\cite{JT} discussed the pion mass, both neutral and charged, in two-flavour
\chpt\ on the lattice. They work in the version of \chpt\ where the fields
satisfy periodic boundary conditions but there are background fields
$\vec B=\vec\theta/L$. They have periodic momenta $\vec p_p=(2\pi\vec n)/L$
and define kinematical momenta $\vec p_k = \vec p_p+\vec B$ which coincide with
our definition $\vec p_{BR}$. However when they define the mass they write
the result in the form\footnote{We have changed their notation and conventions
to make the comparison more clearly.}
\begin{equation}
\label{defmassJT}
m_{JT}^2 = E_0^2-\left(\vec p_p+\vec B+\vec K\right)^2
= E_0^2-\left(\vec p_p+\vec B\right)^2
-2\left(\vec p_p+\vec B\right)\cdot\vec K+\mathrm{NNLO}.
\end{equation}
$\vec K$ is NLO, thus we can neglect $\vec K^2$ as indicated.
Comparing (\ref{defmassBR}) and (\ref{defmassJT}), the
parts containing the integral $A_\mu^V$ in (\ref{resultmasses}) can be written
in the form $-2(\vec p_p+\vec B)\cdot\vec K$. 
\cite{JT} expresses this that the meson field (spatial) momentum is
renormalized. When comparing the expressions, keep in mind we have
also a twist on the sea quarks while \cite{JT} does not.

Comparing with the results of \cite{twisted} is not obvious. The masses
are not defined there. The discussion of loop diagrams in the main text
indicates that they used momenta of the form $\vec p_p+\vec B$ everywhere
and if one assumes that their mass is defined as
\begin{equation}
m_{SV1}^2 = E_0^2-\left(\vec p_p+\vec B\right)^2\,,
\end{equation}
then they missed the terms with $A_\mu^V$. If instead a definition of the mass
similar to (\ref{defmassJT}) is assumed we are in agreement.
The expression
corresponding to $\vec K$ is not present in \cite{twisted}.

For the decay constants a similar issue arises. They are not fully defined
in \cite{twisted}. If one defines the decay constant from the time component
of the axial current then only the parts $\Delta^V F_M$ are relevant and we are
in full agreement, if, as is natural, the neutral pion and eta
decay constants in \cite{twisted} are defined with the isospin and octet
axial currents.
It turns out that to NLO the decay constants
can be defined with a shift in momentum $\vec K^\prime$ similar to what was done
for the masses, i.e. the full matrix element has the form
\begin{equation}
\label{defFpi2}
\left< 0 |A_\mu^M | M(p) \right> = i\sqrt{2}F_M \left(p_\mu+K_\mu^\prime\right)
+ \mathrm{NNLO}\,.
\end{equation}
However, the needed shift vector is different in the two cases,
\begin{equation}
\vec K \ne \vec K^\prime\,.
\end{equation}

The pion form-factors as discussed in Sect.~\ref{sec:formfactor} were treated
in the two-flavour case in \cite{JT}. They discussed the time component only
but added partial twisting and quenching.
The extra terms in the matrix element (\ref{deffh}) are seen
in (19) of \cite{JT} as well. The terms in (19) in \cite {JT}
containing $G_{FV},G^{\mathrm{iso}}_{FV},\mathbf{G}^{\mathrm{iso}}_{FV}$
correspond to our $\Delta^V f_+,\Delta^V f_-,\Delta^V h_\mu$ of
(\ref{resultf+}), (\ref{resultf-}) and (\ref{resulth}).
We have included the spatial components as well and checked that the expected
Ward identity following from current conservation is satisfied when all
effects of the boundary condition are taken into account.
It should be noted that here the matrix element cannot be rewritten in
terms of one form-factor $f_+$ and momenta rescaled with a
shift $\vec K^{\prime\prime}$.

\section{Conclusions}
\label{sec:conclusions}

In this paper we discussed the one-loop tadpole and bubble integrals
in finite volume and at non-zero twist. 

We have worked out the expressions in one-loop ChPT for masses, axial-vector
and pseudo-scalar decay
constants as well as the vacuum expectation value and the two-point
function for the electromagnetic current. We also discussed how
the vector form-factors behave at finite twist angle.
In particular we showed how one needs more form-factors than in the
infinite volume limit and obtained expressions for those at one-loop order.
We discussed how the extra terms are needed in order for the Ward identities
to be satisfied.

Explicit formulas are provided for a large number of cases. We have given
numerical results for all masses and the axial-vector decay
constant of the charged mesons. We found that for the vector form-factor there
are nontrivial finite volume effects due to the extra form-factors and have
discussed the size of these effects on the form-factors. In particular, we
have taken care to precisely define what all quantities are.

Work is in progress for including the effects due
to partial quenching and twisting as well as the effects from
staggered fermions \cite{future}.

\acknowledgments
This work is supported, in part, by the European Community SP4-Capacities
``Study of Strongly Interacting Matter'' (HadronPhysics3, Grant Agreement number 283286) and
the Swedish Research Council grants 621-2011-5080 and 621-2013-4287.

\appendix
\section{Finite volume integrals with twist}
\label{app:Integrals}

The basic method to do finite volume integrals with twist can be found in
\cite{twisted}. The discussion below follows \cite{sunsetfiniteV} closely.

\subsection{Miscellaneous formulae}
\label{app:formulae}

The first ingredient is the Poisson summation formula
which is in one dimension
\begin{align}
\label{eq:Poisson}
\frac{1}{L}
\sum_{\begin{smallmatrix}k = 2\pi n/L + \theta/L\\n\in\mathbb{Z}
\end{smallmatrix}} f(k) &= 
\sum_{m\in\mathbb{Z}} \int \frac{dk}{2\pi}f(k)e^{iLmk}e^{-im\theta}.
\end{align}
The $\sum_{m\in\mathbb{Z}}e^{im a}$ projects on 
$a=2\pi n$. $k-\theta/L$ is of this form, hence the sign in
$e^{-im\theta}$ in (\ref{eq:Poisson}).

The results for loop integrals with twist are expressed with
the third Jacobi theta function
and its derivatives w.r.t. to $u$.
The definitions are
\begin{align}
\label{eq:Theta_3}
\Theta_3(u,q) &= \sum_{n=-\infty}^\infty q^{n^2}e^{2\pi iun}\,,
\quad
\Theta_3^\prime(u,q) = \sum_{n=-\infty}^\infty q^{n^2}2\pi ine^{2\pi iun},
\nonumber\\
\Theta_3^{\prime\prime}(u,q) &= -\sum_{n=-\infty}^\infty q^{n^2}4\pi^2n^2e^{2\pi iun}.
\end{align}
Some useful properties can be found in \cite{sunsetfiniteV}.


\subsection{Tadpole integral}

We define the tadpole integral in finite volume with twist as
\begin{align}
  A^{\{~,\mu,\mu\nu\}}(m^2_M,n) &= \frac{1}{i}\int_V \frac{d^d k}{(2\pi)^d} 
\frac{\{1,k^\mu,k^\mu k^\nu\}}{(k^2-m^2_M)^n}\,.
\end{align}
The blank in the superscript indicates no superscript.
$\int_V d^d k/(2\pi)^d$ is defined in (\ref{defintV}). 
The momentum $\vec k$ which is summed over must be such that
the boundary condition for the propagating meson $M$ is satisfied,
\begin{align}
\label{eq:k}
\vec k &= \frac{2\pi}{L}\vec n+\frac{\vec\theta_M}{L}\,,
\quad \vec\theta_M=(\theta_M^x,\theta_M^y,\theta^z_M)\,.
\end{align}
We also introduce a fourvector $\theta_M=(0,\vec\theta)$.
Note that this implies that the tadpole integral
is not invariant under $\vec k\to -\vec k$ since $-\vec k$ does not satisfy
the boundary conditions for non-zero twist. The direction of propagation is
important. We drop the subscript $M$ below for clarity.

To describe the evaluation of these integrals, we restrict to the case
$\{1\}$ and then quote the results for the other cases. We Wick rotate to
Euclidean space and apply Poisson's summation formula
from Eq. (\ref{eq:Poisson}), giving
\begin{align}
  A(m^2,n) &= (-1)^n\sum_{\vec{l}\in\mathbb{Z}^3}\int \frac{d^d k_E}{(2\pi)^d} \frac{1}{(k_E^2+m^2)^n}e^{iL\vec{l}\cdot\vec{k}-i\vec{l}\cdot\vec{\theta}}\,.
\end{align}
The term with $\vec{l} = 0$ gives the infinite volume
result. We focus on the finite volume part and use a prime on the sum
to indicate that we sum over $\vec{l} \neq 0$. Using
$1/a^n = (1/\Gamma(n))\int_0^\infty d\lambda \lambda^{n-1} e^{-a\lambda}$,
we get
\begin{align}
  &A^V(m^2,n) = (-1)^n\sum^\prime_{\vec{l}\in\mathbb{Z}^3}
\int \frac{d^d k_E}{(2\pi)^d}\int \frac{d\lambda}{\Gamma(n)}\lambda^{n-1}
e^{-\lambda(k^2+m^2)} e^{iL\vec{l}\cdot\vec{k}-i\vec{l}\cdot\vec{\theta}}.
\end{align}
The shift of integration variable via $k = \bar k+iLl/(2\lambda)$,
with $l=(0,\vec l)$, completes the square:
\begin{align}
    A^{V}(m^2,n) &= (-1)^n\sum^\prime_{\vec{l}\in\mathbb{Z}^3}
\int \frac{d^d \bar k_E}{(2\pi)^d}\int \frac{d\lambda}{\Gamma(n)}\lambda^{n-1}e^{-\lambda(\bar k^2+m^2)} e^{-L^2{\vec{l}}^2/(4\lambda)-i\vec{l}\cdot\vec{\theta}}.
\end{align}
We can now perform the Gaussian integral and we end up with
\begin{align}
  &A^{V}(m^2,n) = (-1)^n\sum^\prime_{\vec{l}\in\mathbb{Z}^3}
\int \frac{d\lambda}{\Gamma(n)}\frac{\lambda^{n-1-d/2}}{(4\pi)^{d/2}}
 e^{-\lambda m^2} e^{- L^2{\vec{l}}^2/(4\lambda)-i\vec{l}\cdot\vec{\theta}}.
\end{align}
Changing variables $\lambda\rightarrow \lambda L^2/4$ and using the Jacobi
theta function of (\ref{eq:Theta_3}), we arrive at
\begin{align}
A^{V}(m^2,n) &= (-1)^n\left(\frac{L^2}{4}\right)^{n-2}\int \frac{d\lambda}{\Gamma(n)}\frac{\lambda^{n-3}}{(4\pi)^{2}} e^{-\lambda m^2L^2/4}
\left(\prod_{j=x,y,z} \Theta_3\left(\frac{-\theta^j}{2\pi},e^{-1/\lambda}\right)-1\right).
\end{align}
The $-1$ removes the case with $\vec l=0$ and the triple product comes
from the triple sum and we set $d=4$.

Performing the same operations using the other elements in $X$ gives
for the finite volume corrections
\begin{align}
  A^{V\mu}(m^2,n) &=
 (-1)^{n}\frac{1}{\pi L}\left(\frac{L^2}{4}\right)^{n-2}
\int \frac{d\lambda}{\Gamma(n)}\frac{\lambda^{n-4}}{(4\pi)^2}
 e^{-\lambda m^2L^2/4}
\nonumber\\  &\times
 \Theta_3^\prime\left(\frac{-\theta^\mu}{2\pi},e^{-1/\lambda}\right)
\prod_{\begin{smallmatrix}j=x,y,z\\j\ne\mu
  \end{smallmatrix}}
 \Theta_3\left(\frac{-\theta^j}{2\pi},e^{-1/\lambda}\right).
\end{align}
Note that the component $\mu=0$ vanishes.
\begin{align}
A^{V\mu\nu}(m^2,n) &= g^{\mu\nu}A^V_{22}(m^2,n)+A^{V\mu\nu}_{23}(m^2,n)\,,
\nonumber\\
A^V_{22}(m^2,n)&= \frac{(-1)^{n-1}}{2}\left(\frac{L^2}{4}\right)^{n-3}
\int \frac{d\lambda}{\Gamma(n)}\frac{\lambda^{n-4}}{(4\pi)^{2}}
e^{-\lambda m^2L^2/4}
 \left(\prod_{j=x,y,z} \Theta_3\left(\frac{-\theta^j}{2\pi},e^{-1/\lambda}\right)
    -1\right),
\nonumber\\
A^{V\mu\nu}_{23}(m^2,n) &=
\frac{(-1)^{n}}{4\pi^2}\left(\frac{L^2}{4}\right)^{n-3}\int \frac{d\lambda}{\Gamma(n)}\frac{\lambda^{n-5}}{(4\pi)^2} e^{-\lambda m^2L^2/4} 
\nonumber\\
  &\hskip-8mm((a) \mu=0\text{ or }\nu=0)~ \times 0
\nonumber\\
  &\hskip-8mm((b)0\ne\mu\ne\nu\ne0) ~\,\times
     \Theta_3^\prime\left(\frac{-\theta^\mu}{2\pi},e^{-1/\lambda}\right)
       \Theta_3^\prime\left(\frac{-\theta^\nu}{2\pi},e^{-1/\lambda}\right)
\!\prod_{\begin{smallmatrix}j=x,y,z\\j\ne\mu,\nu\end{smallmatrix}}\!
  \Theta_3\left(\frac{-\theta^j}{2\pi},e^{-1/\lambda}\right)
 \nonumber\\
  &\hskip-8mm((c)\mu=\nu\ne0) ~~~~~~ \times\Theta_3^{\prime\prime}\left(\frac{-\theta^\mu}{2\pi},e^{-1/\lambda}\right)
\prod_{\begin{smallmatrix}j=x,y,z\\j\ne\mu\end{smallmatrix}}
  \Theta_3\left(\frac{-\theta^j}{2\pi},e^{-1/\lambda}\right)
\end{align}
$A^{V\mu\nu}_{23}$ vanishes for $\mu=0$ or $\nu=0$, case (a).
For $\mu\ne\nu$ one uses the line (b), otherwise (c). $A^{V\mu\nu}_{23}$ is from
the $l^\mu l^\nu$ part after the shift of $k$ to $\bar k$.
The sign conventions are Minkowski with upper indices
as indicated. In the main text we have dropped the argument $n$, we only need
$n=1$.

\subsection{Two propagator integrals}

We define two propagator integrals as
\begin{align}
\label{defB}
  B^{\{~,\mu,\mu\nu\}}(m_1^2,m_2^2,n_1,n_2) &= \frac{1}{i}\int_V \frac{d^d k}{(2\pi)^d} \frac{\{1,k^\mu,k^\mu k^\nu\}}{(k^2-m_1^2)^{n_1}((q-k)^2-m_2^2)^{n_2}}.
\end{align}
As in the tadpole case, the direction of the propagators is important.
We use the convention that the particles propagate in the direction
of the momentum indicated in the propagator. We thus writing $k$ and $q-k$
in the propagators to indicate this, even if the sign in the denominator
at first sight is not relevant. 

We have in principle a twist angle vector for each of the
two particles in the denominators. However, it is sufficient to
specify only the twist vector for the first propagator, with $m_1^2$,
and the external momentum $q$. The latter must be such that $q-k$ automatically
produces the correct boundary conditions for the particle corresponding
to $m_2^2$. This is discussed in detail in \cite{twisted}.

We first do the Poisson summation trick to get full integrals over $k$.
We combine the two propagators in (\ref{defB}) using a Feynman parameter $x$
and shift integration variable by $k=\tilde k+xq$.
We then have expressions of the form of the previous subsection but with
$\tilde k$ as integration variable and
$\tilde{m}^2 = (1-x)m_1^2+xm_2^2-x(1-x)q^2$ instead of $m^2$, as well
as $\vec{\tilde\theta} = \vec\theta_1-x\vec q$.

The final result is
\begin{align}
B^{V}(m_1^2,m_2^2,n_1,n_2,q) &= \frac{\Gamma(n_1+n_2)}{\Gamma(n_1)\Gamma(n_2)}
\int_0^1 dx (1-x)^{n_1-1}x^{n_2-1}A^{V}(\tilde{m}^2,n_1+n_2)\,,
\nonumber\\
B^{V\mu}(m_1^2,m_2^2,n_1,n_2,q) &= 
\frac{\Gamma(n_1+n_2)}{\Gamma(n_1)\Gamma(n_2)}\int_0^1 dx (1-x)^{n_1-1}x^{n_2-1}
\nonumber\\  &\times
 \left( A^{V\mu}(\tilde{m}^2,n_1+n_2) + xq^\mu A^{V}(\tilde{m}^2,n_1+n_2) \right),
\nonumber\\
  B^{V\mu \nu}(m_1^2,m_2^2,n_1,n_2) &= 
 \frac{\Gamma(n_1+n_2)}{\Gamma(n_1)\Gamma(n_2)}\int_0^1 dx (1-x)^{n_1-1}x^{n_2-1}
 \left( A^{V\mu\nu}(\tilde{m}^2,n_1+n_2)
 \right.
\nonumber\\ &
\left.
  + x(q^\mu g_\alpha^\nu+q^\nu g_\alpha^\mu) A^{V\alpha}(\tilde{m}^2,n_1+n_2)
 + x^2q^\mu q^\nu A^{V}(\tilde{m}^2,n_1+n_2)\right)\,.
\end{align}
The signs are for upper indices in Minkowski space as indicated.
For the numerical evaluation it is useful to treat the integral over $x$ and
$\lambda$ together. In the main text we have dropped the indices $n_1$ and
$n_2$ and used the components as defined below in (\ref{defcomponents}).

\subsection{Integral relations}
\label{app:relations}

It is possible to derive relations between integrals
using the relation
\begin{align}
\label{PVtrick}
2 k\cdot q &= (k^2-m_1^2)-((q-k)^2-m_2^2)+m_1^2-m_2^2+q^2\,.
\end{align}
These were done in infinite volume in \cite{PV} and in \cite{ABT}
in the same conventions as ours.
The trick remains valid at finite volume. Care has to be taken in the shift
of integration momentum for some of the tadpole integrals (from $k$ to $q-k$)
but that is consistent with the boundary conditions.

We define components
\begin{align}
\label{defcomponents}
 B^{V\mu}(m_1^2,m_2^2) &= 
 q^\mu B_1^{V}(m_1^2,m_2^2,q) + B_2^{V\mu}(m_1^2,m_2^2,q)
\nonumber\\
  B^{V\mu\nu}(m_1^2,m_2^2,q) &= 
   q^\mu q^\nu B_{21}^{V}(m_1^2,m_2^2,q)
 + g^{\mu\nu}B_{22}^{V}(m_1^2,m_2^2,q) 
 + B^{V\mu\nu}_{23}(m_1^2,m_2^2,q)\,.
\end{align}
The relations we get from using (\ref{PVtrick}) are, suppressing the
arguments $(m_1^2,m_2^2,q)$,
\begin{align}
\label{eq:Brelations}
2 q^2  B^{V}_1 &= -A^V(m_1^2)+A^V(m_2^2) + (q^2+m_1^2-m_2^2)B^V
 - 2 B^{V\mu}_2 q_\mu\,,
\nonumber\\
q_\mu  B_{23}^{V\mu\nu} &= -q^2q^\nu B_{21}^{V} - q^\nu B_{22}^{V} 
\nonumber\\   &
+ \frac{1}{2}\left(- A^{V\nu}(m_2^2) - A^{V\nu}(m_1^2) + q^\nu A(m_2^2) + (q^2 + m_1^2 - m_2^2) B^{V\nu} \right)\,.
\end{align}
These are valid for $n_1=n_2=1$ and $n=1$ in the tadpole integrals.
They are needed to prove the Ward identities in the main text.
We have also used them to simplify the expressions.

\end{document}